\documentclass[journal,compsoc]{IEEEtran}
\usepackage{amsmath}
\usepackage{amssymb}
\usepackage[noend]{algpseudocode}
\usepackage{algorithmicx,algorithm}
\usepackage{diagbox}
\usepackage{caption3}
\usepackage{subfigure} 
\usepackage{multirow}
\usepackage{graphicx}  
\usepackage{color}
\usepackage{xr}
\usepackage{hyperref}
\newtheorem{theorem}{\bf Remark}
\newtheorem{definition}{\bf Definition}

\makeatletter
\newcommand*{\addFileDependency}[1]{
  \typeout{(#1)}
  \@addtofilelist{#1}
  \IfFileExists{#1}{}{\typeout{No file #1.}}
}
\makeatother

\newcommand*{\myexternaldocument}[1]{%
    \externaldocument{#1}%
    \addFileDependency{#1.tex}%
    \addFileDependency{#1.aux}%
}
\myexternaldocument{Appendix}

\ifCLASSOPTIONcompsoc
  \usepackage[nocompress]{cite}
\else
  \usepackage{cite}
\fi
\ifCLASSINFOpdf
\else
\fi
\begin{document}
%
\title{Adaptive 3D Mesh Steganography Based on Feature-Preserving Distortion}
%
%
%
%

\author{
	Yushu~Zhang, \IEEEmembership{Senior Member, IEEE},
	Jiahao~Zhu,
	Mingfu~Xue, \IEEEmembership{Senior Member, IEEE},
	Xinpeng Zhang, \IEEEmembership{Member, IEEE}, and Xiaochun Cao, \IEEEmembership{Senior Member, IEEE}
     \IEEEcompsocitemizethanks{
   \IEEEcompsocthanksitem Yushu Zhang and Mingfu Xue are with the College of Computer Science and Technology, Nanjing University of Aeronautics and Astronautics, Nanjing 211106, China, and also with the Zhengzhou Xinda Institute of Advanced Technology, Zhengzhou 450001, China. E-mail: $\{$yushu, mingfu.xue$\}$@nuaa.edu.cn.
     \IEEEcompsocthanksitem Jiahao Zhu is with the School of Computer Science and Engineering, Sun Yat-sen University, Guangzhou, 510006, China. E-mail: Zhujh59@mail2.sysu.edu.cn.
     \IEEEcompsocthanksitem Xinpeng Zhang is with School of Computer Science, Fudan University, Shanghai 200433, China. E-mail: zhangxinpeng@fudan.edu.cn.
     \IEEEcompsocthanksitem Xiaochun Cao is with School of Cyber Science and Technology, Sun Yat-sen University, Shenzhen, 518107, China. E-mail: caoxiaochun@mail.sysu.edu.cn.}
\thanks{Manuscript received XXX; revised XXX; accepted XXX.}
\thanks{Date of publication XXX; data of current version XXX}
\thanks{This work was supported in part by the National Key R$\&$D Program of China under Grant 2021YFB3100400 and in part by the Open Foundation of Henan Key Laboratory of Cyberspace Situation Awareness under Grant HNTS2022013.}
\thanks{(Corresponding author: Jiahao Zhu.)}
\thanks{Digital Object Identifier no. XXX}
}

%
%

\markboth{Journal of \LaTeX\ Class Files,~Vol.~14, No.~8, August~2015}%
{Shell \MakeLowercase{\textit{et al.}}: Bare Demo of IEEEtran.cls for Computer Society Journals}
%



\IEEEtitleabstractindextext{%
\begin{abstract}Current 3D mesh steganography algorithms relying on geometric modification are prone to detection by steganalyzers. In traditional steganography, adaptive steganography has proven to be an efficient means of enhancing steganography security. Taking inspiration from this, we propose a highly adaptive embedding algorithm, guided by the principle of minimizing a carefully crafted distortion through efficient steganography codes. Specifically, we tailor a payload-limited embedding optimization problem for 3D settings and devise a feature-preserving distortion (FPD) to measure the impact of message embedding. The distortion takes on an additive form and is defined as a weighted difference of the effective steganalytic subfeatures utilized by the current 3D steganalyzers. With practicality in mind, we refine the distortion to enhance robustness and computational efficiency. By minimizing the FPD, our algorithm can preserve mesh features to a considerable extent, including steganalytic and geometric features, while achieving a high embedding capacity. During the practical embedding phase, we employ the $Q$-layered syndrome trellis code (STC). However, calculating the bit modification probability (BMP) for each layer of the $Q$-layered STC, given the variation of $Q$, can be cumbersome. To address this issue, we design a universal and automatic approach for the BMP calculation. The experimental results demonstrate that our algorithm achieves state-of-the-art performance in countering 3D steganalysis. Code is available at \href{https://github.com/zjhJOJO/3D-steganography-based-on-FPD.git}{https://github.com/zjhJOJO/3D-steganography-based-on-FPD.git}.
\end{abstract}

}

\maketitle

\IEEEdisplaynontitleabstractindextext

%
\IEEEpeerreviewmaketitle

\IEEEraisesectionheading{\section{Introduction}\label{sec:introduction}}

%
%
%
%
\IEEEPARstart{S}{teganography} is both an art and a science that conceals confidential messages imperceptibly in innocent-looking digital media, such as images, videos, audio, and texts \cite{2009Steganography}. In recent years, the rapid advancement of 3D technologies has led to the widespread use of 3D models in various applications, such as virtual reality and 3D printing. This trend has made 3D models suitable hosts for steganography, similar to other digital media. Currently, 3D models are typically represented in three different ways: point clouds, polygon meshes, and voxels \cite{Qi_2017_CVPR}. Among these representations, meshes are preferred by 3D steganographers because they provide a greater capacity for message embedding.

In 3D mesh steganography, most algorithms rely on moving mesh vertices to embed messages. Early algorithms \cite{Cayre2003,Wang2005,Cheng2006,Chao2009,Itier2017,Li2017} focused on enlarging the embedding capacity while minimizing mesh distortion. However, with the advent of 3D steganalyzers \cite{LFS208,LFS52,LFS64,LFS76,ELFS124,NVT2021,WFS228}, these algorithms are now easily detectable. Consequently, steganographers have shifted their focus to steganographic security, i.e., anti-steganalysis ability. Li \emph{et al.} \cite{LZY2017} were the first to do so, but their algorithm can resist only a few 3D steganalyzers. Numerous studies in 2D steganography have demonstrated that incorporating adaptability into algorithms is an effective means of enhancing steganographic security. This strategy relies on the principle of minimal impact embedding  \cite{fridrich2007practical}, which involves two core procedures: minimizing a suitably-defined distortion and practical steganography coding. Since Syndrome Trellis Code (STC) \cite{FIller2011} has become the standard methodology for practical embedding, most approaches to adaptive steganography strive to find a distortion model that performs well experimentally. Examples include the classical spatial algorithms HUGO \cite{HUGO2010}, WOW \cite{WOW2012}, and S-UNIWARD \cite{S-UNIWARD2014}, as well as some algorithms tailored for JPEG images \cite{S-UNIWARD2014, Guo2014, Guo2015}. Building on this idea, Zhou \emph{et al.} \cite{VND2018} applied the design framework for 2D adaptive steganography \cite{Gibbs} to 3D settings. They proposed two geometric-feature-related distortion models: vertex normal based distortion and curvature based distortion. Although their algorithm achieves satisfactory results in countering steganalysis, it involves nonadaptive embedding, namely least significant bit replacement (LSBR), which somewhat limits its potential for improving steganographic security.

From the above short review, it is clear that the distortion model plays a crucial role in achieving effective concealment of messages. A well-designed distortion can capture specific characteristics that impede detection by steganalyzers within a chosen feature space \cite{VND2018}. Building on this insight, we propose a highly adaptive 3D mesh steganography method based on feature-preserving distortion (FPD). Our algorithm comprises three key steps: embedding domain construction, distortion design, and practical message embedding. For the first step, we begin with a Decimal-to-Integer (D2I) conversion for vertex coordinates, and then use the integers for constructing an embedding domain composed of bitplanes. In the second step, we propose an additive distortion model FPD that measures the impact of message embedding with the assistance of 3D steganalytic features. Specifically, we evaluate the steganalysis ability of each subfeature from the state-of-the-art statistical 3D steganalyzer, and then, we select the most effective ones for designing the FPD. Furthermore, to make our algorithm practical, we improve upon FPD from the aspect of subfeature selection and computational efficiency. By minimizing the FPD, we can obtain the optimal vertex-changing distribution, which serves as a guide for the subsequent practical embedding. In the last step, we adopt the $Q$-layered STC for bit embedding due to its near-optimal performance for arbitrary additive distortion functions \cite{FIller2011}. However, varying $Q$ makes the calculation of the bit modification probability (BMP) in each layer of $Q$-layered STC troublesome. To address this problem, we devise a universal and automatic BMP calculation approach, dubbed U$\&$A-BMP. The contributions of our work are fourfold:
\begin{itemize}
	\item We present a highly adaptive 3D mesh steganography algorithm that balances security and embedding capacity. It assigns a specific cost to each vertex coordinate change, enabling it to carefully consider the direction and distance of each vertex movement. To our knowledge, this is the first 3D steganography to achieve such a high level of adaptability.
	\item We assess the effectiveness of steganalytic features adopted by the state-of-the-art 3D steganalyzer and create the FPD to gauge the impact of message embedding. Our experimental results indicate that FPD significantly enhances the security of our approach while retaining mesh features, such as steganalytic and geometric features, to a certain extent.
	\item We fine-tune the FPD design, resulting in a significant enhancement of its robustness and computational efficiency. This improvement makes our algorithm practical and feasible for real-world scenarios.
	\item We provide a toolbox, U$\&$A-BMP, which offers a more convenient and flexible method to calculate the BMP for each layer of $Q$-layered STC, especially when $Q$ varies.
\end{itemize}

The remainder of the paper is organized as follows. In Section 2, we introduce existing 3D mesh steganography and steganalysis. In Section 3, we first describe the overall framework of our algorithm and then elaborate on it from three aspects, i.e., embedding domain construction, distortion design, and message embedding and retrieval. In Section 4, we report the experimental results and discuss the limitations of our algorithm. Finally, we conclude our work in Section 5 and offer several future research topics.
\section{Related Work}
\subsection{3D Mesh Steganography}
Unlike digital watermarking, steganography was developed for covert communication. Existing 3D mesh steganography can be roughly divided into the following four categories \cite{ZhouS}.

\textbf{Two-state steganography}. Cayre and Macq \cite{Cayre2003} proposed a substitutive blind 3D mesh steganographic scheme in the spatial domain, the main idea of which is to regard a triangular surface as a two-state geometrical object. During data embedding, the states of triangles are geometrically modulated in terms of the message to hide. To enlarge the embedding capacity, Wang and Cheng \cite{Wang2005} proposed a multi-level embedding procedure (MLEP). They built a triangle neighbor table and heuristically constructed a $k$-D tree to improve efficiency. Subsequently, they proposed a modified MLEP and combined the spatial domain with representation domain for a larger embedding capacity \cite{Cheng2006}. Later, Chao \emph{et al.} \cite{Chao2009} proposed a novel multilayered embedding scheme, which increases the embedding capacity up to $21\!\sim\!39$ bits per vertex (bpv). Unlike the prior work, Itier \emph{et al.} \cite{Itier2017} provided a novel scheme that synchronizes vertices along a Hamiltonian path and displaces them with static arithmetic coding. Recently, Li \emph{et al.} \cite{Li2017} developed an embedding strategy that allows us to adjust embedding distortion below a speciﬁed level by defining a truncated space. With the emergence of modern 3D steganalyzers, Li \emph{et al.} \cite{LZY2017} were the first to consider the anti-steganalysis ability in 3D mesh steganography, and  they reinvestigated and modified the prior work \cite{Itier2017} to make it more resistant against steganalysis.

\textbf{LSB steganography}. Yang \emph{et al.} \cite{linear2012} investigated the correlation between the spatial noise and normal noise in meshes. Additionally, they proposed a high-capacity 3D mesh steganography that calculates an appropriate quantization level for each vertex and replaces unused LSBs with bits to hide. Zhou \emph{et al.} \cite{VND2018} analyzed the steganalytic features proposed in \cite{LFS64} and selected the most effective parts for their distortion model design. In addition, they combined STCs and LSBR for practical message embedding.

\textbf{Permutation steganography}. Some researchers have explored alternative embedding domains to avoid the mesh distortion caused by embedding changes. Bogomjakov \emph{et al.} \cite{Bogomjakov2008} proposed a distortionless permutation steganographic scheme for polygon meshes, which hides messages by permutating the storage order of faces and vertices. Building upon the work of Bogomjakov \emph{et al.}, Huang \emph{et al.} \cite{Huang2009} designed a more efficient version. Moreover,  they conducted a theoretical analysis that indicates that the expected embedding capacity of their algorithm is closer to optimal. Tu \emph{et al.} \cite{Tu2010} further improved upon the work \cite{Bogomjakov2008} by making slight modifications to the mapping strategy. Later, Tu \emph{et al.} \cite{Tu2015} constructed a skewed binary coding tree rather than a complete binary tree for permutation steganography. However, the average embedding capacity of this algorithm is less than that reported in \cite{Tu2010}. Therefore, Tu \emph{et al.} \cite{Tu2012} subsequently built a maximum expected level tree based on a novel message probability model, achieving a higher expected embedding capacity than that of all previous methods.

\textbf{Transform-domain-based steganography}. Aspert \emph{et al.} \cite{Aspert2002} utilized the watermarking technique presented in \cite{Wagner2000} and performed embedding operations by making slight displacements of vertices. Maret \emph{et al.} \cite{Maret2004} improved upon the method developed by Aspert \emph{et al.} by creating a similarity-transform invariant space and by adapting the embedding procedure to the sample distribution in that space. Kaveh \emph{et al.} \cite{Kaveh2015} proposed 3D polygonal mesh steganography by using the surfacelet transform to achieve higher capacity and less geometric distortion.
\begin{figure*}[t]
	\centering
	\includegraphics[width=0.85\linewidth]{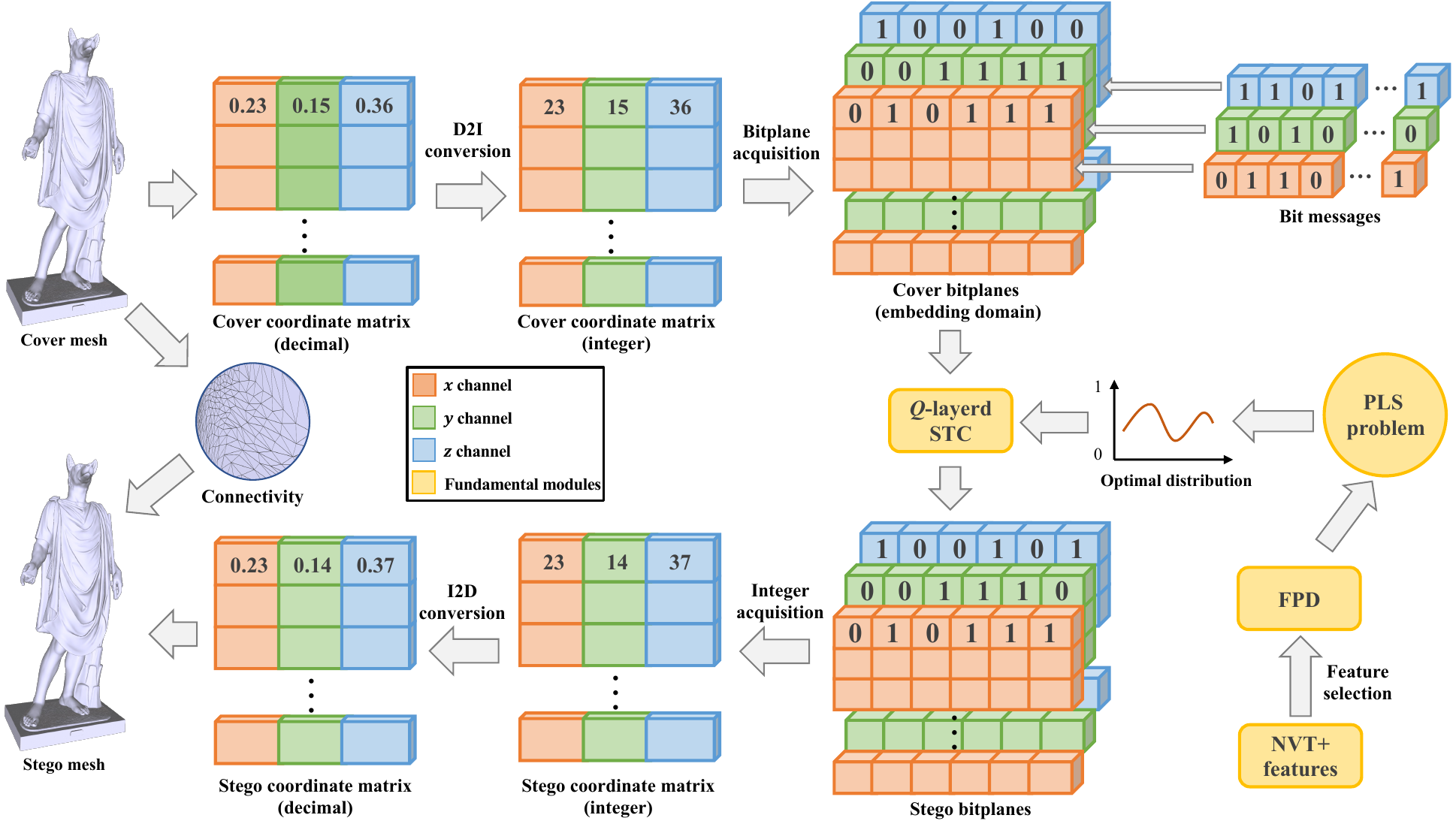}
	\caption{Message embedding pipeline of the proposed adaptive 3D mesh steganography algorithm based on FPD.}
	\label{A1}
\end{figure*}
\subsection{3D Mesh Steganalysis}
Today, the majority of 3D steganalyzers are designed to detect steganography based on imperceptible geometric modifications. These types of steganalyzers can be implemented in three simple steps.

\textbf{Mesh smoothing and normalization}. First, the original mesh undergoes a smoothing procedure (e.g., Laplacian smoothing) to generate its reference counterpart. Subsequently, both meshes will be transformed by principal component analysis (PCA). Finally, the resulting meshes are scaled to fit within a unit cube centered at the point (0.5, 0.5, 0.5).

\textbf{Calibrated feature extraction}. The calibrated steganalytic features are extracted by computing the feature difference between the original mesh and its reference. Importantly, different steganalyzers may extract features in distinct ways. Yang and Ivrissimtzis \cite{LFS208} designed YANG208, the first 3D steganalytic feature set whose components are relevant to vertex coordinates, norms in Cartesian and Laplacian coordinate systems, dihedral angles of edges, and face normals. Li and Bors \cite{LFS52} reduced the dimension of YANG208 and considered three additional features, i.e., vertex normal, Gaussian curvature, and curvature ratio. Eventually, a 52-dimensional feature set, LFS52, was produced for steganalysis. Kim \emph{et al.} \cite{LFS64} considered more geometric features, such as edge normals, average curvature, and total curvature, and they combined these additional features with LFS52 to create a new steganalytic feature set named LFS64. Since the features in LFS64 are not directly related to the embedding domain, it is likely that the classifier  cannot achieve an ideal classification result if these features are directly input. Hence, they add a homogeneous kernel map before  the training step. The concept of utilizing the 3D spherical coordinate system for steganalytic feature extraction was initially suggested in LFS76 \cite{LFS76}, followed by the inclusion of further edge-related geometric features as supplementary features in ELFS124 \cite{ELFS124}. In contrast to the steganalysis algorithms outlined earlier, Zhou \emph{et al.} \cite{NVT2021} posited that message embedding can disrupt the interdependencies between neighboring faces. Based on this, they introduced neighborhood-level representation-guided normal voting tensor (NVT) features for 3D mesh steganalysis and integrated them with LFS64 to create a novel feature set named NVT+. Regarding mesh smoothing and normalization, Li and Bors \cite{8576574} explored how the smoothing coefficient and iteration number of Laplacian smoothing impact steganalysis performance. More recently, they developed a wavelet-based steganalytic feature set, WFS228, for detecting stego meshes \cite{WFS228}, which significantly enhanced their ability to detect meshes generated by wavelet-based watermarking algorithms.

\textbf{Classification}. The extracted steganalytic features are finally fed into a binary classifier. Currently, the Fisher linear discriminator (FLD) ensemble \cite{FLD2012} has been widely adopted in contemporary 3D steganalyzers due to its fast training speed and highly reliable detection accuracy.

Aside from the steganalysis approaches described above, there are two distinct steganalysis algorithms created specifically for detecting PCA-based steganography \cite{VND2018} and permutation steganography \cite{breakingps}.
\section{Proposed Algorithm}
\subsection{Overall Framework}
Our 3D mesh steganography follows the principle of minimal embedding impact \cite{fridrich2007practical} and builds upon the adaptive framework proposed in \cite{Gibbs}. The message embedding pipeline is shown in Fig. \ref{A1}. Specifically, given a mesh, we first perform the D2I conversion on the vertex coordinates for the embedding domain construction (see Section 3.3). Second, we design the FPD with features selected from the state-of-the-art 3D steganalytic feature set NVT+ \cite{NVT2021}. The specific form of the FPD and its advanced version are described in Section 3.4 and Section 3.5, respectively. To obtain the optimal vertex-changing distribution, we formulate a payload limited sender (PLS) problem tailored for 3D mesh steganography (see Section 3.2) with the FPD. By using the optimal distribution obtained, we employ the $Q$-layered STC to embed the message bits into the cover bitplanes (see Section 3.6), producing the stego bitplanes. Subsequently, the stego bitplanes undergo a Integer-to-Decimal (I2D) conversion and are combined with the topological information of the cover mesh to produce the stego mesh. The meshes used in this work are all composed of triangular faces unless otherwise stated. 
\subsection{Construction of the PLS Problem}
Let $\boldsymbol{M} = \{\boldsymbol{V},\boldsymbol{E},\boldsymbol{F}\}$ be a cover mesh with vertex set  $\boldsymbol{V}=\{\boldsymbol{v}_{i}\}_{i=1}^{|\boldsymbol{V}|}$, where $\boldsymbol{v}_{i}=[v_{ix},v_{iy},v_{iz}]^{\top}$, edge set $\boldsymbol{E} =\{\boldsymbol{e}_i\}_{i=1}^{|\boldsymbol{E}|}$, and face set $\boldsymbol{F}=\{\boldsymbol{f}_i\}_{i=1}^{|\boldsymbol{F}|}$. The impact of message embedding is measured by a distortion function $D$. For the sake of security, the sender should embed messages while minimizing $D$. In this work, $D$ is limited to the following additive form,
\begin{equation}\label{fpd}
	D(\boldsymbol{M},\boldsymbol{M}')=\sum_{j\in \{x,y,z\}}D_j(\boldsymbol{M},\boldsymbol{M}')=\sum_{j\in \{x,y,z\}}\sum_{i=1}^{|\boldsymbol{V}|}\rho(\delta_{ij}).
\end{equation}
 Here $\boldsymbol{M}'=\{\boldsymbol{V}',\boldsymbol{E}',\boldsymbol{F}'\}$ is the stego counterpart of $\boldsymbol{M}$, where $\boldsymbol{V'}=\{\boldsymbol{v'}_{i}\}_{i=1}^{|\boldsymbol{V'}|}$ and $\boldsymbol{v'}_{i}=[v'_{ix},v'_{iy},v'_{iz}]^{\top}$;  and $\delta_{ij}=v'_{ij}-v_{ij}$ denotes a change quantity; $\rho: \mathbb{R}\rightarrow[0,\infty)$ defines a cost function that measures the cost of  $\delta_{ij}$ occurring. Eq. (\ref{fpd}) indicates that $\rho(\delta_{ij})$ is \emph{independent of changes made at other vertices and coordinate channels, implying that the embedding changes do not interact}. Eq. (\ref{fpd}) also shows the preliminary form of the FPD, and obviously, the effectiveness for our algorithm relies on the design of $\rho$. 
 
As mentioned previously, the embedding changes are noninteracting. Therefore, the most straightforward approach to perform message embedding is to decompose the entire process into three subtasks according to the number of coordinate channels. For each channel, we hide a fixed average payload of $m_j$ bits, where $j\in\{x,y,z\}$. Considering that $\boldsymbol{M}$ is predetermined and our algorithm does not alter its topology (implying $\boldsymbol{E}=\boldsymbol{E}'$ and $\boldsymbol{F} = \boldsymbol{F}'$),  we can abbreviate $D(\boldsymbol{M},\boldsymbol{M}')$ as $D(\boldsymbol{V}')$. Consequently, the PLS problem for each channel can be defined as 
\begin{equation}\label{PLS}
\mathop{\rm{minimize}}_{p}\ E_{p}[D_j(\boldsymbol{V}')] \quad
\mathop{\rm{s.t.}} H(p)=m_j,
\end{equation}
where $p$ denotes the vertex-changing distribution that we want to calculate and $H$ is the information entropy function. The specific form and solution method of the PLS problem (\ref{PLS}) are discussed in the next section.

\subsection{Construction of Embedding Domain}

Most of the meshes used in prior works are sourced from public 3D datasets, and their vertex coordinates are typically represented in decimal form, such as $\boldsymbol{v}_i=[0.172345,0.267324,0.563521]^\top$, with each coordinate value reserving a limited number of significant digits. As a result, it is always possible to find the minimum natural number $k_{ij}$ that satisfies $v_{ij}\times 10^{k_{ij}} \in \mathbb{Z}$. We then define $k_{d}=max(k_{ij})$, and the D2I conversion shown in Fig. {\ref{A1}} can be achieved by $\ddot{v}_{ij} = v_{ij}\times10^{k_d}$.

Likewise, we can easily determine the minimum positive integer $h_b$ satisfying $0 \leq \ddot{v}_{ij} \leq 2^{h_b}-1$ for any $\ddot{v}_{ij}$. By letting $b_{ij}^{(l)}$ be the $l$-th least significant bit of $\ddot{v}_{ij}$, we can construct a bitplane $\boldsymbol{B}_{j}^{(l)}=\{b_{ij}^{(l)}\}_{i=1}^{|\boldsymbol{V}|}$. Then the embedding domain can be defined as $\{\boldsymbol{B}_j^{(l)}|j\in\{x,y,z\},l\in\{1,\cdots,h_b\}\}$.  
 
As our message embedding involves modifying bitplanes, which is equivalent to adding an integer vector to $\ddot{\boldsymbol{v}}_i$, we can generate the final stego vertices using the I2D conversion, 
\begin{equation}\label{dm}
	\boldsymbol{v}'_i = \begin{bmatrix}
		v'_{ix} \\
		v'_{iy}  \\
		v'_{iz} 
	\end{bmatrix} = \begin{bmatrix}
		(\ddot{v}_{ix} + \lfloor\delta_{ix}\times10^{k_d}\rfloor)/10^{k_d}\\
		(\ddot{v}_{iy} + \lfloor\delta_{iy}\times10^{k_d}\rfloor)/10^{k_d}\\
		(\ddot{v}_{iz} + \lfloor\delta_{iz}\times10^{k_d}\rfloor)/10^{k_d}\end{bmatrix}.
\end{equation}
While in theory $\delta_{ij}$ can take any value, in our constructed embedding domain, modifying the bitplanes causes $\ddot{v}_{ij}$ to vary by an integer only. Thus, we round down $\delta_{ij}\times10^{k_d}$, which can also be viewed as discretizing $\delta_{ij}$ directly. We define $\delta_{ij} \in \boldsymbol{I} \subset \{\delta/10^{k_d}|\delta\in \mathbb{Z}\}$, where the cardinality of $\boldsymbol{I}$ is limited to avoid perceptible mesh distortion resulting from message embedding. This enables us to express problem (\ref{PLS}) as follows:
\begin{equation}\label{PLSD}
	\begin{aligned}
		&\mathop{\rm{minimize}}_{p}\ E_p[D_j(\boldsymbol{V}')]=\sum_{i=1}^{|\boldsymbol{V}|}\sum_{\delta_{ij}\in \boldsymbol{I}}\rho(\delta_{ij})p(\delta_{ij})\\
		&\begin{aligned}
			&\mathop{\rm{s.t.}} \ H(p)=\sum_{i=1}^{|\boldsymbol{V}|}\sum_{\delta_{ij}\in \boldsymbol{I}}p(\delta_{ij}){\rm{ln}}\frac{1}{p(\delta_{ij})}=m_j.\\
		\end{aligned}
	\end{aligned}
\end{equation}
The optimal $p$ has a form of Gibbs distribution \cite{Gibbs}, namely
\begin{equation}\label{gibbsf}
	p(\delta_{ij})= \frac{e^{-\lambda_j\rho(\delta _{ij})}}{\sum_{\delta \in \boldsymbol{I}} e^{-\lambda_j\rho(\delta)}}, j \in \{x,y,z\}.\\
\end{equation}
$\lambda_j$ is a scalar calculated through an iterative search with the condition $H(p)=m_j$. Once $\rho$ is determined, we can calculate the optimal $p$ with Eq. (\ref{PLSD}) and Eq. (\ref{gibbsf}).
\subsection{Design of FPD}
\subsubsection{Selection of Efficient Steganalytic Features}\label{3.4.1}
The emergence of a new information security technique is inevitably accompanied by corresponding attacks, and the ensuing back-and-forth between attackers and defenders promotes progress on both sides. It is on this basis that we believe well-designed steganalyzers can contribute to the creation of more secure steganography.

In 3D steganalysis, NVT+, a 100-dimensional feature set $\boldsymbol{\Phi}=\{\phi_{i}\}^{100}_{i=1}$, has been demonstrated experimentally to possess the best steganalysis performance \cite{NVT2021,ZhouS}. Although the idea of preserving all steganalytic features during message embedding may be enticing, it is challenging to achieve in reality. Therefore, we need to identify which subfeatures of NVT+ are effective for steganalysis. The evaluation of subfeatures in NVT+ is presented in Appendix A, and we conclude that subfeatures $\phi_{1} \!\sim \!\phi_{12}$, $\phi_{33} \!\sim \!\phi_{36}$, $\phi_{65} \!\sim \!\phi_{76}$, $\phi_{77} \!\sim \!\phi_{88}$, and $\phi_{89} \! \sim \!\phi_{100}$ may be useful for designing a distortion model.
\subsubsection{Design of Cost Function}
Without loss of generality, we choose the first $L$ best subfeatures from NVT+ to construct the cost function $\rho$ as below. 
\begin{equation}\label{CF}
	\rho(\delta_{ij})=\mu\sum_{k=1}^{L}N_{or}(||S_k(\boldsymbol{M})-S_k(\boldsymbol{M}'(\delta_{ij}))||_1)
\end{equation}
where $S_{k}$ denotes an extractor w.r.t. the $k$-th subfeature, outputting a feature vector; $\boldsymbol{M}'(\delta_{ij})$ denotes that only $v_{ij}$ in $\boldsymbol{M}$ is altered and its change quantity is $\delta_{ij}$; $N_{or}$ is a normalization function; $\mu > 0$ is a scaling factor set to prevent numerical calculation issues. Given that the steganalytic subfeatures mentioned above originate from different geometric features, and their values may be of diverse magnitude orders, $N_{or}$ plays a crucial role in balancing their contribution to $\rho$. Substituting Eq. ({\ref{CF}}) into Eq. (\ref{fpd}) provides us with the specific form of the FPD. By minimizing the FPD, steganalytic, local, and global geometric features can be preserved to a certain extent, which is well-demonstrated in subsequent experiments.
\subsection{Improvement to FPD}
Although subfeatures $\phi_{1} \!\sim \!\phi_{12}$, $\phi_{33} \!\sim \!\phi_{36}$, $\phi_{65} \!\sim \!\phi_{76}$, $\phi_{77} \!\sim \!\phi_{88}$ and $\phi_{89} \! \sim \!\phi_{100}$ exhibit satisfactory performance in steganalysis,  their suitability for our distortion model requires further verification.

\begin{figure}[!t]
	\centering
	\subfigure[$N_1(\boldsymbol{v}_i)$]{\label{fig2:a}\includegraphics[width=0.45\linewidth]{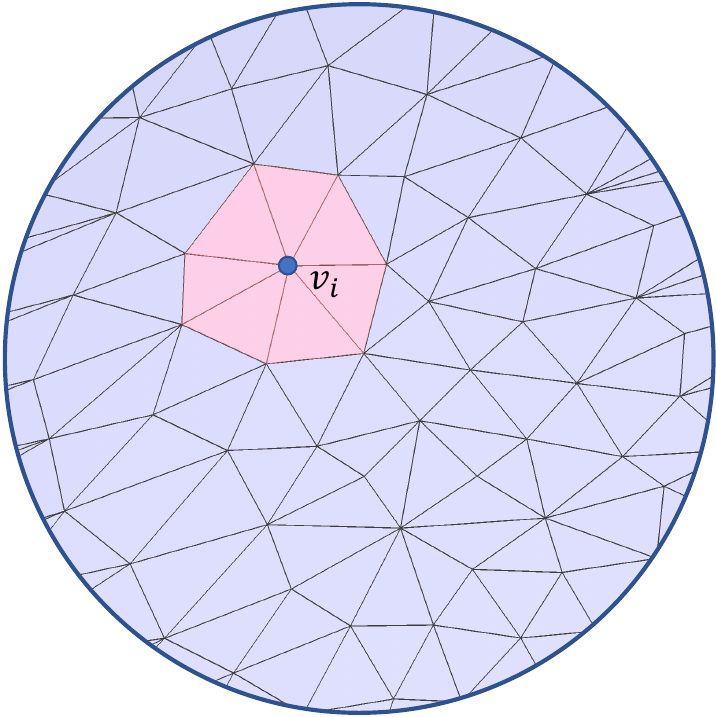}}
	\subfigure[$ID_{1}(\boldsymbol{v}_i)$]{\label{fig2:b}\includegraphics[width=0.45\linewidth]{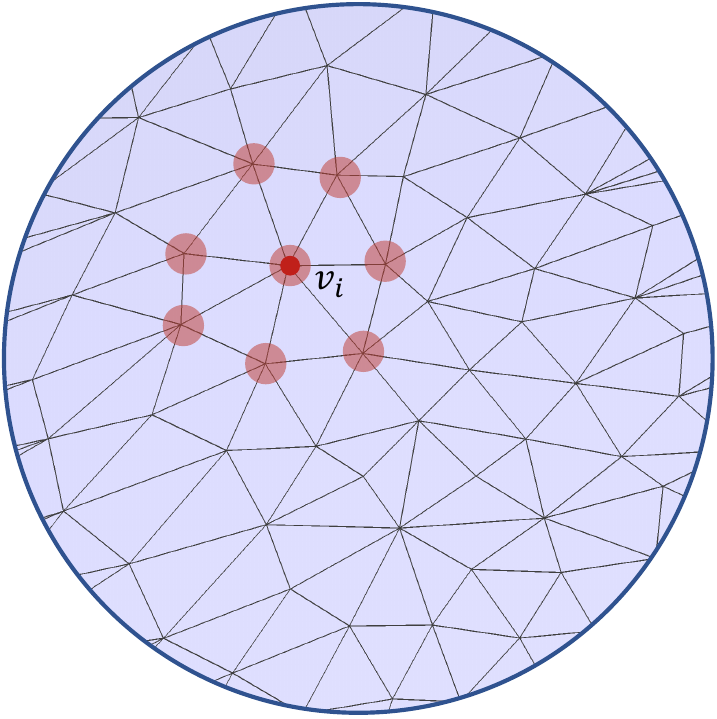}} \\
	\subfigure[$N_2(\boldsymbol{f}_i)$]{\label{fig2:c}\includegraphics[width=0.45\linewidth]{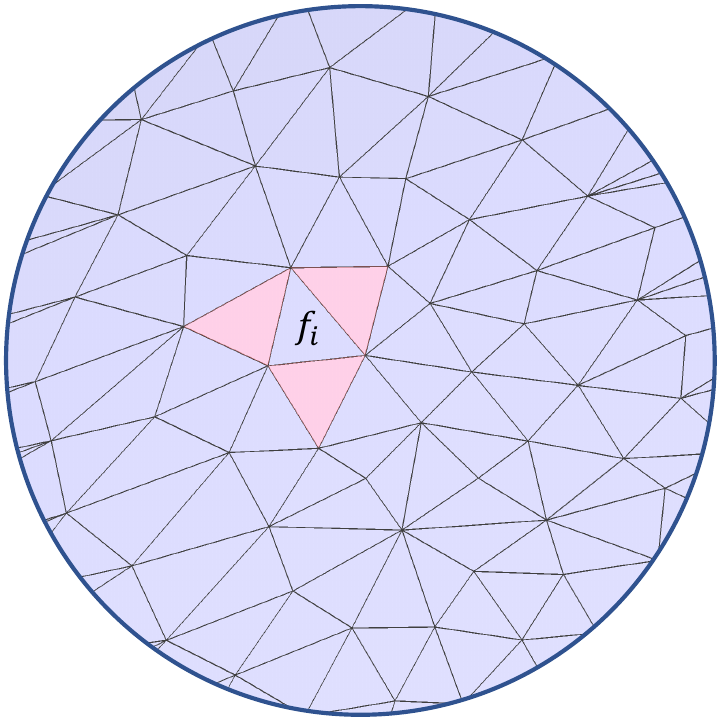}}
	\subfigure[$ID_{2}(\boldsymbol{v}_i)$]{\label{fig2:d}\includegraphics[width=0.45\linewidth]{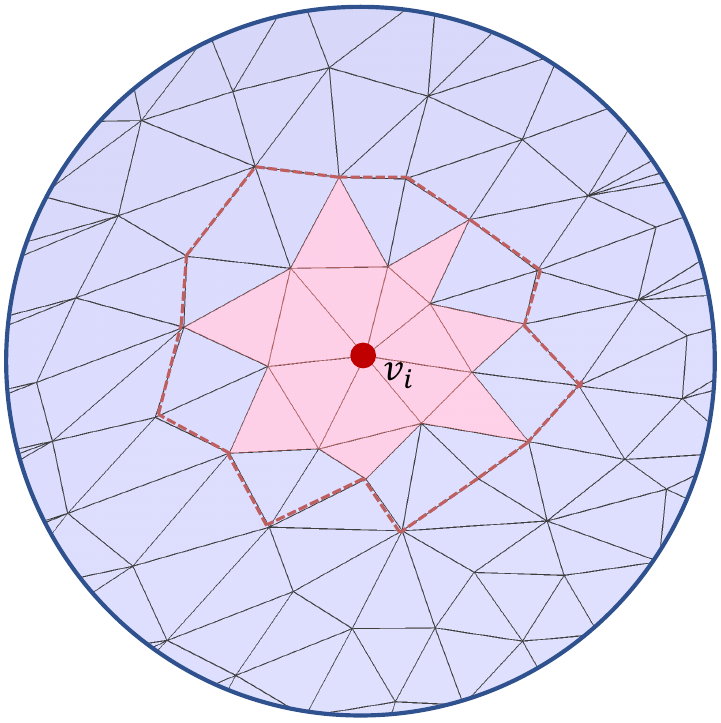}}\\
	\subfigure[$N_3(\boldsymbol{f}_i)$]{\label{fig2:e}\includegraphics[width=0.45\linewidth]{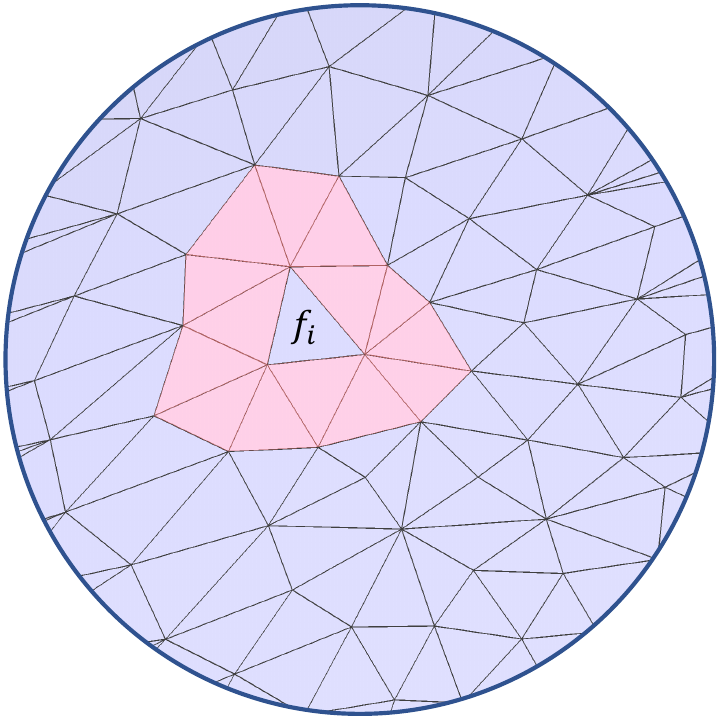}}
	\subfigure[$ID_{3}(\boldsymbol{v}_i)$]{\label{fig2:f}\includegraphics[width=0.45\linewidth]{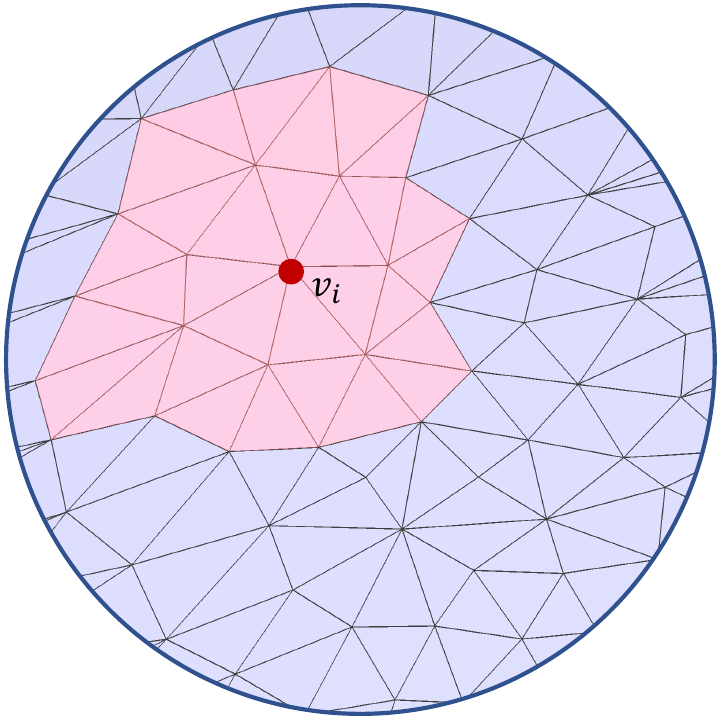}}
	\caption{Illustration of three neighborhood patterns and their respective vertex influence regions. (a) $N_1(\boldsymbol{v}_i)$ is a set of 1-ring neighboring faces of $\boldsymbol{v}_i$. (c) $N_2(\boldsymbol{f}_i)$ a set of faces sharing edges with $\boldsymbol{f}_i$. (e) $N_3(\boldsymbol{f}_i)$ is a set of faces sharing vertices with $\boldsymbol{f}_i$. (b), (d), and (f) are the visualizations of the vertex influence domains corresponding to (a), (c), and (e), respectively. The modified vertices in (b), (d), and (f) are marked in red, and their corresponding vertex influence domains are painted pink.}\label{nh}
\end{figure}
The specific analysis of $\phi_{1} \!\sim \!\phi_{12}$ and $\phi_{33} \!\sim \!\phi_{36}$ is presented in Appendix B. The final result indicates that both of these subfeatures exhibit a lack of generalization against certain specifically designed steganography. Additionally, the utilization of $\phi_{33} \!\sim \!\phi_{36}$ may result in $\rho\equiv0$ for certain vertex movements. Thus, we decide not to consider them in $\rho$. The remaining three subfeatures do not expose the same defects as the first two. In this regard, it is reasonable to include them in Equation (\ref{CF}). However, directly employing the original subfeature extractors from \cite{NVT2021} results in time-consuming calculations of $\rho(\delta_{ij})$ for all vertices and $\delta_{ij} \in \boldsymbol{I}$, as presented in Table \ref{tab2}. To alleviate this problem, we propose an accelerated method as follows.
 \begin{figure*}[t]
	\centering
	\subfigure[]{\label{fig3:a}\includegraphics[width=0.30\linewidth, height=0.32\textwidth]{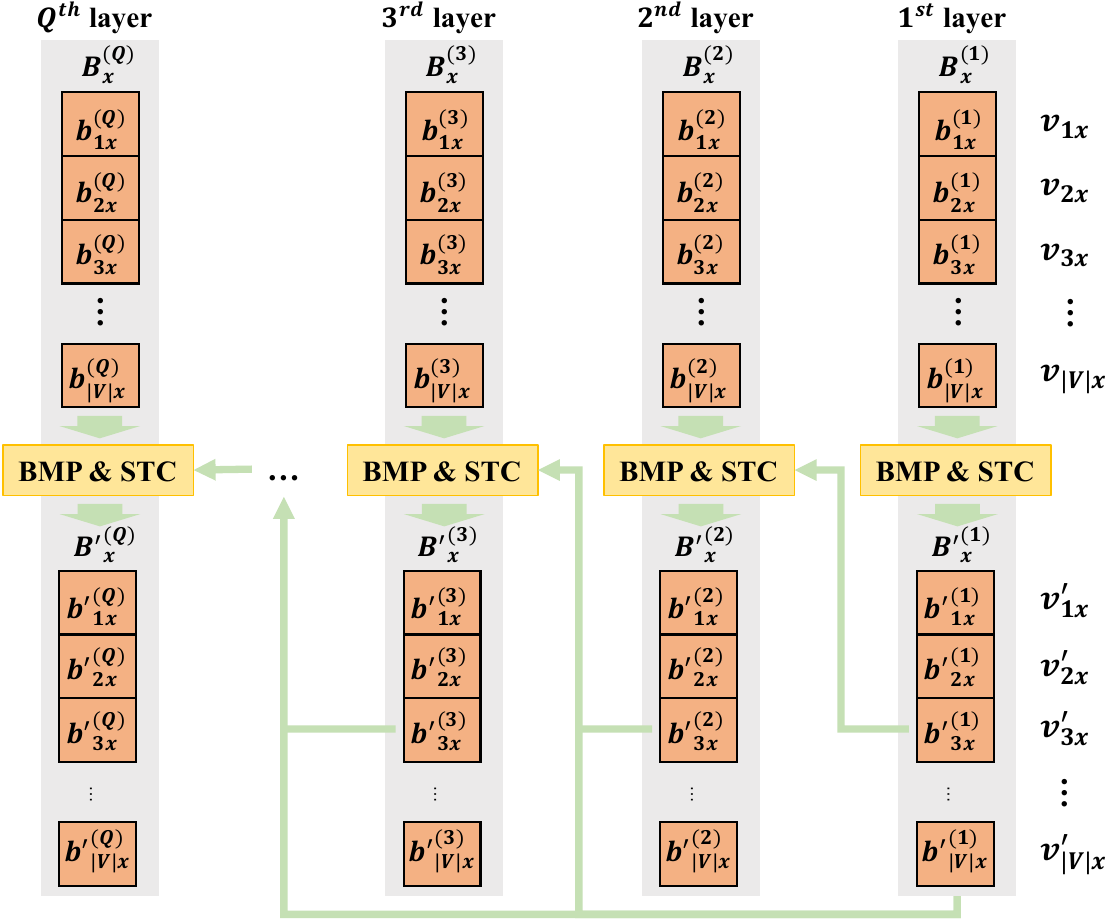}} 
	\subfigure[]{\label{fig3:b}\includegraphics[width=0.66\linewidth]{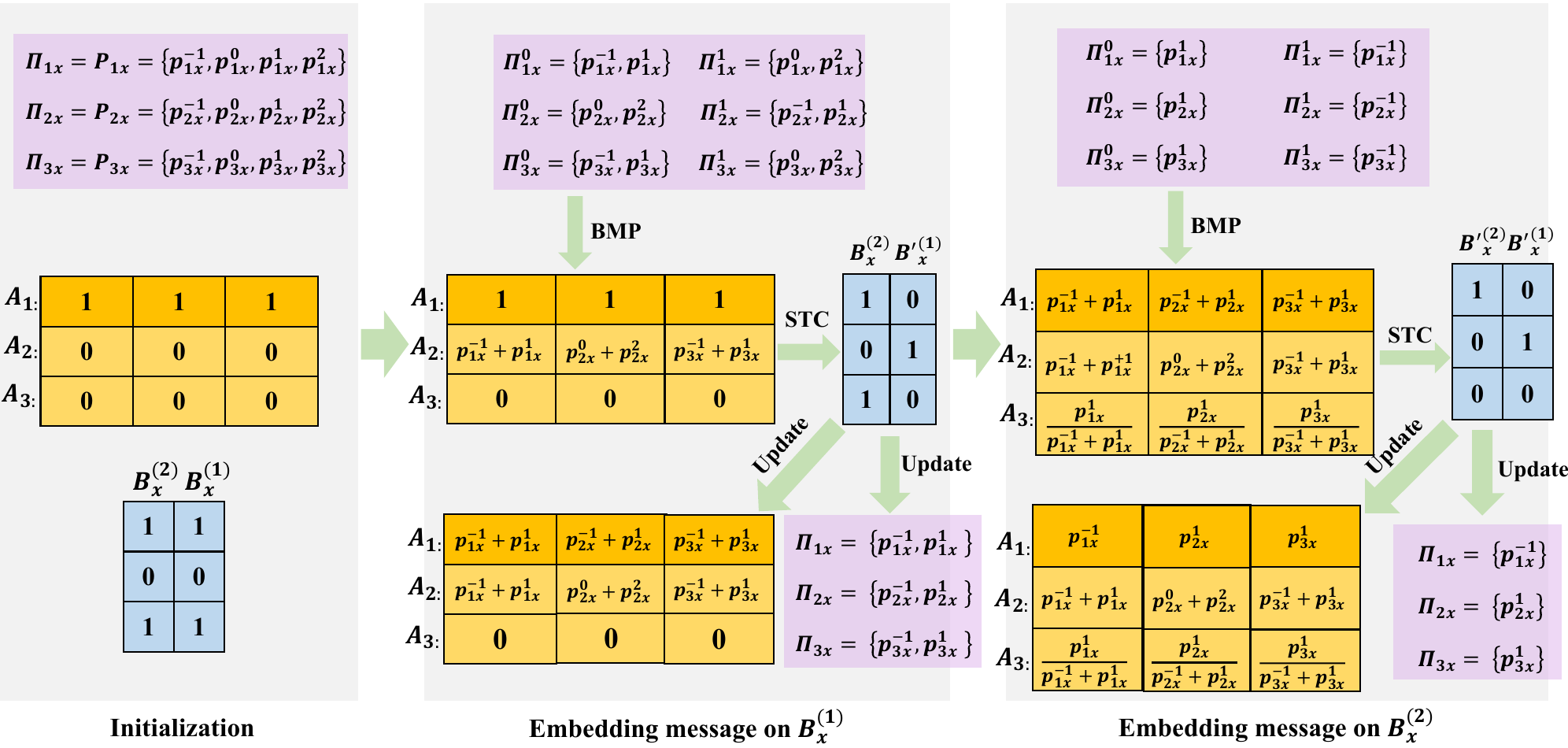}}
	\caption{(a) Basic flow chart of the $Q$-layered STC. (b) A visual example of U$\&$A BMP with $\boldsymbol{I}=\{-1/10^{k_d},0,1/10^{k_d},2/10^{k_d}\}$ and $\boldsymbol{B}_{x}=\{\boldsymbol{B}_x^{(1)},\boldsymbol{B}_x^{(2)}\}$, where $\boldsymbol{B}^{(1)}_x=\{1,0,1\}$ and $\boldsymbol{B}_x^{(2)}=\{1,0,1\}$. Please note that $p^{\delta}_{ij}$ is short for $p(\delta_{ij}=\delta)$, and $\boldsymbol{A}$, except for its first row, records the BMP corresponding to $b'^{(l)}_{ij}=0$ only.}
	\label{psp}
\end{figure*}

Before going any further, we need a brief understanding of the original subfeature extractors w.r.t. $\phi_{65} \!\sim \!\phi_{76}$, $\phi_{77} \!\sim \!\phi_{88}$, and $\phi_{89} \! \sim \!\phi_{100}$. As described in \cite{NVT2021}, the three subfeature extractors rely heavily on their respective predefined neighborhood patterns, i.e., $N_1$, $N_2$, and $N_3$, which are shown in Fig. \ref{fig2:a}, \ref{fig2:c}, and \ref{fig2:e}, respectively. Correspondingly, three types of NVTs of size $3 \times 3$ are defined by
\begin{equation}\label{nvt}
	\begin{aligned}
		&\boldsymbol{{{\rm T}}}_{1i}=\sum_{\boldsymbol{f}\in N_1(\boldsymbol{v}_i)} 
		W_1(\boldsymbol{f})  n({\boldsymbol{f}})\cdot n({\boldsymbol{f}})^\top, i\in\{1,\cdots,|\boldsymbol{V}|\},\\
		&\boldsymbol{{\rm T}}_{2i}=\sum_{\boldsymbol{f}\in N_2(\boldsymbol{f}_i)} 
		W_2(\boldsymbol{f}) n({\boldsymbol{f}})\cdot n({\boldsymbol{f}})^\top, i\in\{1,\cdots,|\boldsymbol{F}|\},\\
		&\boldsymbol{{\rm T}}_{3i}=\sum_{\boldsymbol{f}\in N_3(\boldsymbol{f}_i)} 
		W_3(\boldsymbol{f}) n({\boldsymbol{f}})\cdot n({\boldsymbol{f}})^\top, i\in\{1,\cdots,|\boldsymbol{F}|\},
	\end{aligned}
\end{equation} 
where $n({\boldsymbol{f}})$ denotes the face normal of $\boldsymbol{f}$ and the complete form of the weighting coefficient $W_k(\boldsymbol{f})$ is given in Appendix C. Calculating $\rho$ with the original subfeature extractors is also introduced in Appendix C. Here we skip directly to the description of our accelerated method.

Eq. (\ref{nvt}) implies that \emph{each vertex or face of a mesh actually has a corresponding NVT}. Let $\boldsymbol{\rm{T}}_{1i}$, $\boldsymbol{\rm{T}}_{2i}$, and $\boldsymbol{\rm{T}}_{3i}$ be associated with $\boldsymbol{v}_i$, $\boldsymbol{f}_i$, and $\boldsymbol{f}_i$, respectively. After establishing these associations, we introduce a new concept called \emph{vertex influence domain}.
\begin{definition}[Vertex influence domain]\label{d1} Given a neighborhood pattern $N_k$, the influence domain of $\boldsymbol{v}_i$, denoted by $ID_{k}(\boldsymbol{v}_i)$, is a set of vertices or faces, whose corresponding tensors $\boldsymbol{{\rm T}}_{ki}$ are changed when $\boldsymbol{v}_i$ is moved.
\end{definition}
The mathematical expression of $ID_{k}(\boldsymbol{v}_i)$ is given as follows:
\begin{equation}\label{VID}
	\begin{aligned}
		&ID_{1}(\boldsymbol{v}_i)=\{\boldsymbol{v}\mid \boldsymbol{v} \in \boldsymbol{f}, \boldsymbol{f}\in N_1(\boldsymbol{v}_i)\},\\
		&ID_{2}(\boldsymbol{v}_i)=\{\boldsymbol{f}\mid \boldsymbol{f} \in N_2(\boldsymbol{f}^\star), \boldsymbol{f}^\star \in N_1(\boldsymbol{v}_i)\},\\
		&ID_{3}(\boldsymbol{v}_i)=\{\boldsymbol{f}\mid \boldsymbol{f} \in N_3(\boldsymbol{f}^\star), \boldsymbol{f}^\star \in N_1(\boldsymbol{v}_i)\}.
	\end{aligned}
\end{equation}
The visualizations of $ID_{1}(\boldsymbol{v}_i)$, $ID_{2}(\boldsymbol{v}_i)$, and $ID_{3}(\boldsymbol{v}_i)$ are shown in Fig. \ref{fig2:b}, \ref{fig2:d}, and \ref{fig2:f}, respectively.
\begin{theorem}\label{p2}
	Given a watertight mesh $\boldsymbol{M}$ with many vertices and faces and, assuming the neighbor number of each vertex in $\boldsymbol{M}$ is roughly the same, we have $|ID_{1}(\boldsymbol{v}_i)|\ll |\boldsymbol{V}|$, $|ID_{2}(\boldsymbol{v}_i)|\ll |\boldsymbol{F}|$, and $|ID_{3}(\boldsymbol{v}_i)|\ll |\boldsymbol{F}|$.
\end{theorem}
A short explanation of Remark \ref{p2} is given in Appendix D. In reality, meshes satisfying the assumptions in Remark \ref{p2} are not uncommon, e.g. the classical Stanford bunny meshes. 

Making full use of Remark \ref{p2} and $ID_{k}(\boldsymbol{v}_i)$ can markedly speed up the calculation of $\rho(\delta_{ij})$ for each vertex in $\boldsymbol{M}$ and $\delta_{ij}\in\boldsymbol{I}$. This is because when only $\boldsymbol{v}_i$ in $\boldsymbol{M}$ is moved by $\delta_{ij}$, all we need to do is recalculate the eigenvalues of a tiny portion of NVTs that correspond to elements in $ID_{k}(\boldsymbol{v}_i)$ to obtain $S_k(\boldsymbol{M}'(\delta_{ij}))$. The thorough explanation and relevant pseudo code are provided in Appendix E.
\subsection{Implementation of Message Embedding and Retrieval}
\subsubsection{Calculation of BMP}
Prior to introducing our U$\&$A-BMP, it is essential to provide a brief review of the pipeline of $Q$-layered STC. As illustrated in Fig. \ref{fig3:a}, the process of generating stego bitplanes involves calculating BMP and using (1-layered) STC in each layer. Additionally, it is crucial to note that the BMP in the layer $l$ is directly determined by the actual embedding results of all previous $l-1$ layers.  


However, there are some problems with directly applying STC to 3D steganography. First, \cite{FIller2011} provides a manual BMP calculation approach only, and only source code tailored for binary, ternary, and pentary embedding examples are given in their code repository$\footnote{\label{f1}http://dde.binghamton.edu/download/syndrome/}$, which are far from meeting the demand for embedding capacity in 3D settings; Second, the BMP calculation becomes enormously cumbersome as we have to rework our program when $Q$ varies. 
\begin{algorithm}[t]
	\caption{U$\&$A-BMP} 
	\hspace*{0.02in} {\bf Input:} 
	The cover bitplane set $\boldsymbol{B}_j=\{\boldsymbol{B}_{j}^{(l)}\}_{l=1}^{h_b}$, the perturbation set $\boldsymbol{I}$, the optimal distribution set $\boldsymbol{P}_j=\{\boldsymbol{P}_{ij}\}_{i=1}^{|\boldsymbol{V}|}$, where $\boldsymbol{P}_{ij}=\{p(\delta_{ij})\mid \delta_{ij} \in \boldsymbol{I}\}$, $j\in\{x,y,z\}$.\\
	\hspace*{0.02in} {\bf Output:} 
Stego bitplane set $\boldsymbol{B}'_j$.
	\begin{algorithmic}[1]
		\State Set $Q=\inf\mathop{\arg\max}_{q \in \boldsymbol{\mathbb{N}}^*}[|\boldsymbol{I}| \leq 2^q]$ $/\!/$ [$\cdot$] denotes Iverson bracket;
		\State Extend $\boldsymbol{I}$ until $|\boldsymbol{I}|=2^{Q}$, with adding elements' corresponding probability set to 0;
		\State Create an array $\boldsymbol{A}$ of size $(Q+1)\times|\boldsymbol{V}|$ with all entries being 0;
		\State Initialize $\boldsymbol{\Pi}_{:j}=\boldsymbol{P}_{:j}$ and $\boldsymbol{A}_{1:}=1$; $/\!/$``:” means to take all indices in the current dimension.
		\For{$l=2$ to $Q+1$}
		\For{$i=1$ to $|\boldsymbol{V}|$}
		\State $\boldsymbol{\Pi}_{ij}^0=\{p({\delta_{ij}})|p(\delta_{ij}) \in \boldsymbol{\Pi}_{ij},C^{(l-1)}(\ddot{v}_{ij}+\delta_{ij}\times10^{k_d})=0\}$; 
		\State $\boldsymbol{\Pi}_{ij}^{1}=\boldsymbol{\Pi}_{ij}\setminus\boldsymbol{\Pi}_{ij}^{0}$;
		\State $P({b'}_{ij}^{(l-1)}\!=\!0,{b'}_{ij}^{(l-2)},\cdots,{b'}_{ij}^{(1)}) = SUM(\boldsymbol{\Pi}_{ij}^0)$;
		\State $\boldsymbol{A}_{li}=P({b'}_{ij}^{(l-1)}=0,{b'}_{ij}^{(l-1)},\cdots,{b'}_{ij}^{(1)})/\boldsymbol{A}_{1i}$;
		\EndFor
		\State Perform 1-layered STC on $\boldsymbol{B}_j^{(l-1)}$ to generate ${\boldsymbol{B}'}_j^{(l-1)}$;
		\For{$i=1$ to $|\boldsymbol{V}|$}
		\State $\boldsymbol{\Pi}_{ij}=\boldsymbol{\Pi}_{ij}^0$ if ${b'}_{ij}^{(l-1)}=0$, otherwise $\boldsymbol{\Pi}_{ij}=\boldsymbol{\Pi}_{ij}^1$;
		\State $\boldsymbol{A}_{1i}=P({b'}_{ij}^{(l-1)}\mid{b'}_{ij}^{(l-2)},\cdots,{b'}_{ij}^{(1)})\times A_{1i}$;
		\EndFor
		\EndFor
		\State $\boldsymbol{B}'_j=\{\boldsymbol{B}^{(h_b)}_{j},\cdots,\boldsymbol{B}^{(Q+1)}_{j},\boldsymbol{B}'^{(Q)}_{j},\cdots,\boldsymbol{B}'^{(1)}_{j}\}$;
		\State \Return $\boldsymbol{B}'_j$.
	\end{algorithmic}
\end{algorithm}

To facilitate the application of STC in 3D mesh steganography, we design the U$\&$A-BMP, the core of which is described in Algorithm \ref{A1}. Within this algorithm, we employ an array $\boldsymbol{A}$ (excluding its first row) to keep track of the BMP of each bit layer (Line 10) and use the check function $C^{(l)}$ to represent the $l$-th least significant bit of an integer. The first row of $\boldsymbol{A}$ is reserved for storing joint probability. On Lines 9 and 10, ${b'}_{ij}^{(l-2)},\cdots,{b'}_{ij}^{(2)}$, and ${b'}_{ij}^{(1)}$ are used to represent the actual embedding results of previous $l-2$ layers, with their values being omitted for simplicity. Furthermore, the 1-layered STC utilized on Line 11 follows the flipping lemma outlined in \cite{FIller2011}. To explain Algorithm \ref{A1} more clearly, we present an example visualizing the BMP calculation for 2-layered STC in Fig. \ref{fig3:b}. The example begins with variable initialization in Lines 1 - 4, followed by the successive embedding of messages on $\boldsymbol{B}_x^{(1)}$ and $\boldsymbol{B}_x^{(2)}$ by using the for-loop present in Lines 5 - 14.


Notably, Algorithm \ref{A1} is just a pseudo code for the purpose of articulating the logic behind U$\&$A BMP. During practical code implementation, we leverage some algorithmic acceleration techniques to ensure our U$\&$A BMP is computationally efficient, such as the usage of vectorization to avoid an abuse of for-loops.
\subsubsection{Message Embedding}
We present the complete message embedding algorithm in Algorithm 2. We set $|\boldsymbol{I}|=2^Q$ and $\rho(\delta_{ij})<\infty$ for $\delta_{ij} \in \boldsymbol{I}$ to avoid \emph{wetness} in each layer of $Q$-layered STC as much as possible. For more details about wet points in STC, we suggest \cite{FIller2011}.
\begin{algorithm}[t]
	\caption{Adaptive 3D Mesh Steganography Based on FPD} 
	\hspace*{0.02in} {\bf Input:} 
	The cover mesh $\boldsymbol{M}=\{\boldsymbol{V},\boldsymbol{E},\boldsymbol{F}\}$, the quantization parameter $k_d$, the bit number $h_b$, the perturbation set $\boldsymbol{I}$, and the payload $\boldsymbol{m}$.\\
	\hspace*{0.02in} {\bf Output:} 
	Stego mesh $\boldsymbol{M}'$.
	\begin{algorithmic}[1]
		\State Set $m_x=m_y=m_z=\frac{|\boldsymbol{m}|}{3}$;
		\For{$j$ in $\{x,y,z\}$ }
		\State Perform the D2I conversion for all vertices of $\boldsymbol{M}$ and obtain bitplane set $\boldsymbol{B}_j=\{\boldsymbol{B}_{j}^{(l)}\}_{l=1}^{h_b}$;
		\State Calculate $\rho(\delta_{ij})$ for all vertices of $\boldsymbol{M}$ and $\delta_{ij}\in\boldsymbol{I}$ with Eq. (\ref{CF});
		\State Substitute $m_j$ and $\rho(\delta_{ij})$ into Eq. (\ref{PLSD}) and obtain the optimal distribution $\boldsymbol{P}_j=\{\boldsymbol{P}_{ij}\}_{i=1}^{|\boldsymbol{V}|}$, where $\boldsymbol{P}_{ij}=\{p(\delta_{ij})\mid \delta_{ij} \in \boldsymbol{I}\}$;
		\State Input $\boldsymbol{B}_j$, $\boldsymbol{I}$, and $\boldsymbol{P}_j$ to the U$\&$A BMP algorithm and obtain the stego bitplane set $\boldsymbol{B}'_j$;
		\State Perform the I2D conversion on $\boldsymbol{B}'_j$ to generate $\boldsymbol{V}'_j$;
		\EndFor
		\State Concatenate $\boldsymbol{V}'_x$, $\boldsymbol{V}'_y$, and $\boldsymbol{V}'_z$ as $\boldsymbol{V}'$;
		\State $\boldsymbol{M}'=\{\boldsymbol{V}',\boldsymbol{E},\boldsymbol{F}\}$;
		\State \Return $\boldsymbol{M}'$.
	\end{algorithmic}
\end{algorithm}
\subsubsection{Message Retrieval}
Since we adopt the $Q$-layered STC, the whole embedding process can be considered as performing the 1-layered STC on a collection of bitplanes. Accordingly, the message retrieval on each stego bitplane can be implemented by
\begin{equation}
\boldsymbol{m}^{(l)}_j=\boldsymbol{\rm{H}}^{STC} \cdot {\boldsymbol{B}'}_j^{(l)}, l\in \{1,2,\cdots,Q\},\ j \in \{x,y,z\},
\end{equation}
where $\boldsymbol{\rm{H}}^{STC} \in \{0,1\}^{|\boldsymbol{m}^{(l)}_j|\times|\boldsymbol{V}|}$ denotes the parity-check matrix derived from a predesigned submatrix $\widehat{\boldsymbol{\rm{H}}}\in \{0,1\}^{h \times w}$. Typically, $6\leq h\leq15$ and $\frac{1}{w+1}<\frac{|\boldsymbol{m}^{(l)}_j|}{|\boldsymbol{V}|}<\frac{1}{w}$. Note that $\boldsymbol{\rm{H}}^{STC}$is shared between the sender and recipient; the height of $\widehat{\boldsymbol{\rm{H}}}$ is suggested to be set to a moderately large value according to \cite{FIller2011}.
\section{Experimental Results and Analysis}
\subsection{Setup of Experiments}
\subsubsection{Datasets}. All experiments in this paper are conducted on the following three 3D mesh datasets: 
\begin{itemize}
	\item \textbf{Princeton ModelNet (PMN)}$\footnote{http://modelnet.cs.princeton.edu}$: The dataset comprises an extensive assortment of manually-drawn 3D CAD models that cater to researchers in the fields of computer graphics, computer vision, robotics, and cognitive science.
	\item  \textbf{Princeton Shape Benchmark (PSB)}$\footnote{https://segeval.cs.princeton.edu}$: The dataset consists of 400 watertight mesh models, gathered from various web repositories and divided into 20 object categories, with each category comprising 20 meshes.
	\item \textbf{The Stanford Models (TSM)}$\footnote{http://graphics.stanford.edu/data/3Dscanrep/}$: The dataset comprises multiple dense reconstructed mesh models, which were scanned by laser triangulation range scanners. 
\end{itemize}
\begin{table*}[!t]
	\centering
	\caption{Time taken to calculate $\rho(\delta_{ij})$ with six types of cost functions for all vertices of meshes from the PSB, PMN, and TSM datasets and $\delta_{ij}\in\boldsymbol{I}$ when $\boldsymbol{I}\in\{-6/10^{k_d},\cdots,-1/10^{k_d},0,1/10^{k_d},\cdots,6/10^{k_d}\}$.}
	\label{tab2}
	\begin{tabular}{cl|cc|cc|cc}\hline
		& Time (s)&OFPD-S1 & IFPD-S1 & OFPD-S2 & IFPD-S2 & OFPD-S3 & IFPD-S3 \\\hline
		&102.off (15724 vertices) & $\gg 3600$    & \textbf{414.0169}      &  $\gg 3600$     & \textbf{813.1334}      &  $\gg 3600$      & \textbf{761.1621}      \\
		PSB    &140.off (9200 vertices) &  $\gg 3600$      & \textbf{153.5684}     &  $\gg 3600$      & \textbf{286.1365}       &  $\gg 3600$       & \textbf{276.4585}       \\
		&325.off (14999 vertices &  $\gg 3600$     & \textbf{337.3610}    &  $\gg 3600$      & \textbf{739.2899}       &  $\gg 3600$      & \textbf{700.6807}    \\
		\hline
		&1575.off (2397 vertices) &  $2332.8141$    & \textbf{15.9289}      &  $899.4178$     & \textbf{14.5522}      &  1043.5057      & \textbf{11.6842}      \\
		PMN    &305.off (4048 vertices) &  $>3600$      & \textbf{34.7324}      &   2443.0105     & \textbf{31.9157}       &  2361.4446       & \textbf{20.8605}       \\
		&1598.off (5187 vertices)&  $>3600$     & \textbf{49.4106}    &  $>3600$      & \textbf{26.9861}       &  $>3600$       & \textbf{15.7638}    \\\hline
		&bun-zipper-res2.ply (8171 vertices) &  $\gg 3600$    & \textbf{125.1004}      &  $\gg 3600$     & \textbf{227.9974}      &  $\gg 3600$      & \textbf{222.8744}      \\
		TSM   &dragon-vrip-res3.ply (22998 vertices) & $\gg 3600$      & \textbf{888.2709}      &  $\gg 3600$      & \textbf{1794.0436}       &  $\gg 3600$      & \textbf{1735.9160}       \\
		&happy-vrip-res4.ply (7108 vertices) &  $\gg 3600$    & \textbf{99.7105}    &  $\gg 3600$      & \textbf{191.2417}       &  $\gg 3600$       & \textbf{187.3738}    \\ \hline 
	\end{tabular}
\end{table*}
\begin{figure*}[t]
	\centering
	\subfigure[PSB dataset]{\label{fig4:a}\includegraphics[width=0.4\linewidth]{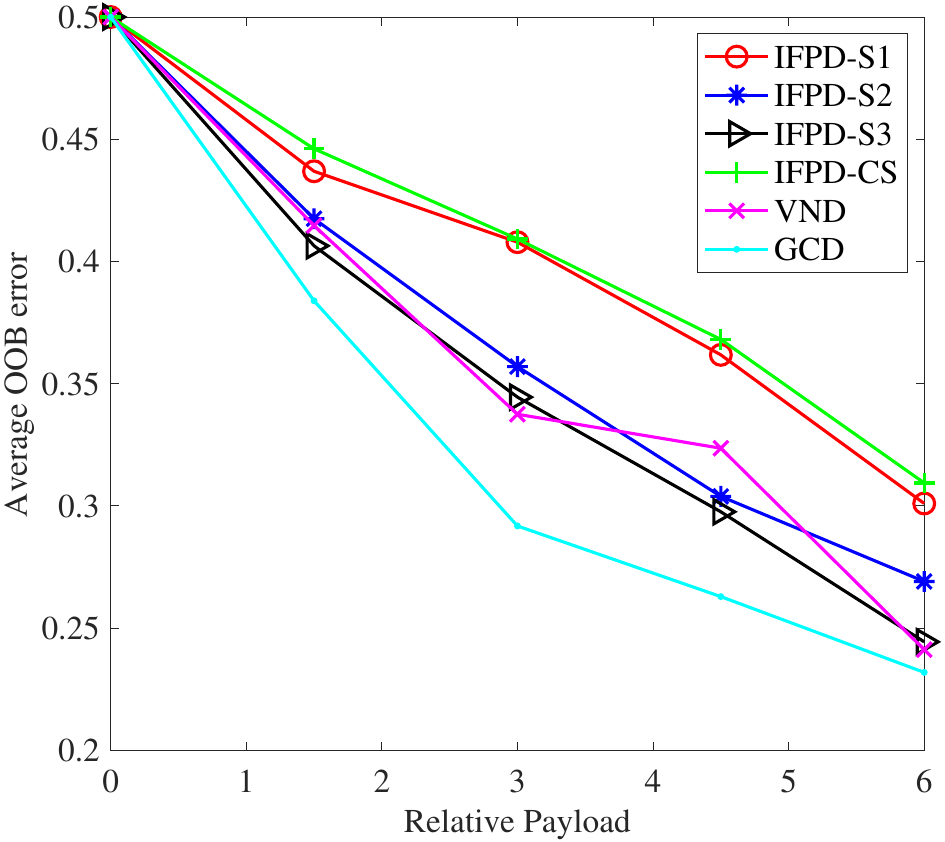}}
	\hspace{2em}
	\subfigure[PMN dataset]{\label{fig4:b}\includegraphics[width=0.4\linewidth]{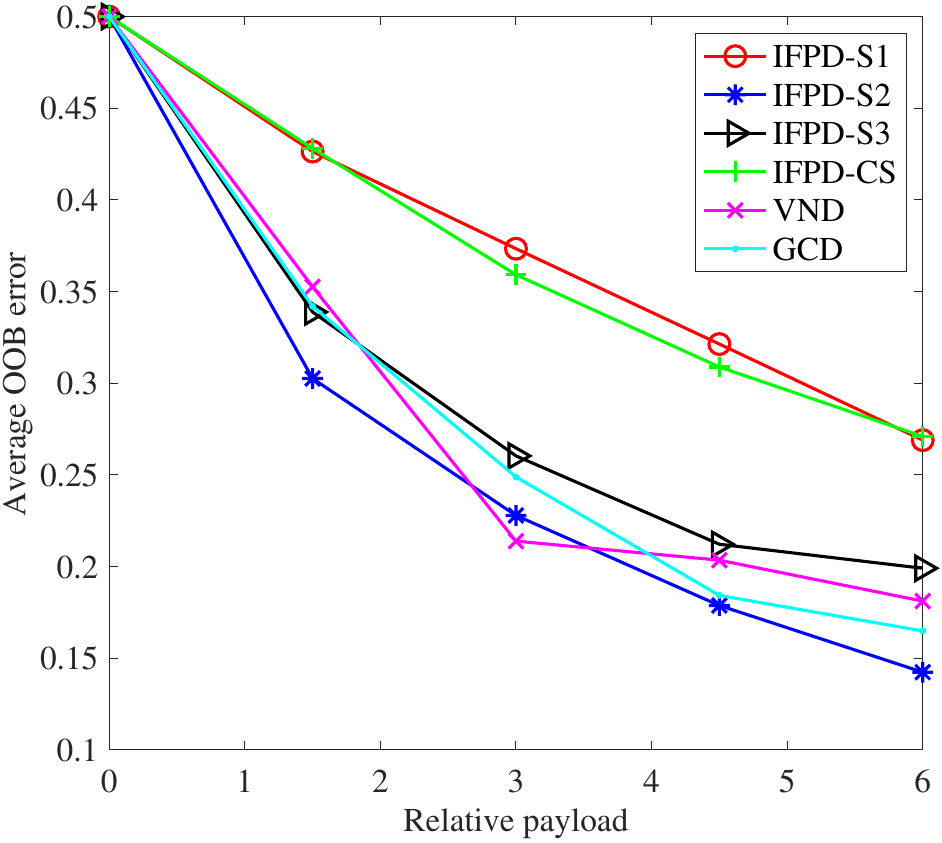}}
	\caption{Average OOB error $\bar{E}_{OOB}$ as a function of the relative payload for six types of distortion functions on (a) the PSB dataset (b) and PMN dataset.} 
	\label{e1}
\end{figure*}
\subsubsection{Criteria for Steganography}
We evaluate our algorithm from three aspects:
\begin{itemize}
	\item\textbf{Steganographic Security}: Steganography security can be understood as the ability of steganographic algorithms to resist steganalysis. In this paper, we utilize the ``out-of-bag" error estimate $E_{OOB}$ \cite{FLD2012} to measure steganography security. It is an unbiased estimate of the testing error, which is defined by
	\begin{equation}
		E^{(n)}_{OOB}=\frac{1}{2N^{data}}\sum_{i=1}^{N^{data}}(B^{(n)}(\boldsymbol{\Phi}_i)+1-B^{(n)}(\boldsymbol{\Phi}'_i)),
	\end{equation}
	where $\boldsymbol{\Phi}_i$ and $\boldsymbol{\Phi}'_i$ are cover and stego steganalytic features respectively; $n$ denotes the number of base learners used in the ensemble classifier, and the base learner is $B^{(n)}(\cdot) \in \{0,1\}$.
	\item\textbf{Embedding Capacity}: Embedding capacity in this paper is measured by the relative payload $\alpha=\frac{m}{|\boldsymbol{V}|}$ bpv (bits per vertex), where $m$ is the size of the practical payload. 
	\item\textbf{Robustness}: The ability to resist various digital attacks, such as affine transformation, vertex
reordering, noise addition, smoothing and simplification.
\end{itemize}
\subsubsection{Experimental Setting}
All experiments are conducted in a computer equipped with an Intel Xeon 5218R processor, 64 GB of RAM, and MATLAB 2019(a).
\subsection{Evaluation of FPD}
\subsubsection{Time Complexity of Cost Calculation} 
Based on the analysis in Section 3.4 and Section 3.5, we have identified that subfeatures $\phi_{65} \sim \phi_{76}$, $\phi_{77} \sim \phi_{88}$, and $\phi_{89} \sim \phi_{100}$ are the most appropriate for the FPD design. However, computing $\rho(\delta_{ij})$ for all vertices and $\delta_{ij} \in \boldsymbol{I}$ with the original method in \cite{NVT2021} is excessively time-consuming. In Section 3.5, we further analyze the three subfeatures and propose an improved version called IFPD. We use the notations OFPD-S1, OFPD-S2, and OFPD-S3 to refer to the three original cost calculation methods based on $\phi_{65} \sim \phi_{76}$, $\phi_{77} \sim \phi_{88}$, and $\phi_{89} \sim \phi_{100}$, respectively. Correspondingly, we denote the improved methods as IFPD-S1, IFPD-S2, and IFPD-S3. To showcase the effectiveness of the IFPD series, we randomly select several meshes from the PSB, PMN, and TSM datasets, and we calculate the vertex-changing cost by using various cost calculation methods with $\boldsymbol{I}=\{\frac{-6}{10^{k_d}},\cdots,\frac{-1}{10^{k_d}},0,\frac{1}{10^{k_d}},\cdots,\frac{6}{10^{k_d}}\}$. Table \ref{tab2} presents the time consumed by each method on each mesh. It is evident from the table that the OFPD-based methods are excessively time-consuming, as processing meshes with more than 7000 vertices can take more than an hour. In contrast, the IFPD series significantly accelerates the cost calculation process.
\begin{figure*}
	\centering
	\subfigure[LFS52, PSB dataset] 
	{\includegraphics[width=4.2cm]{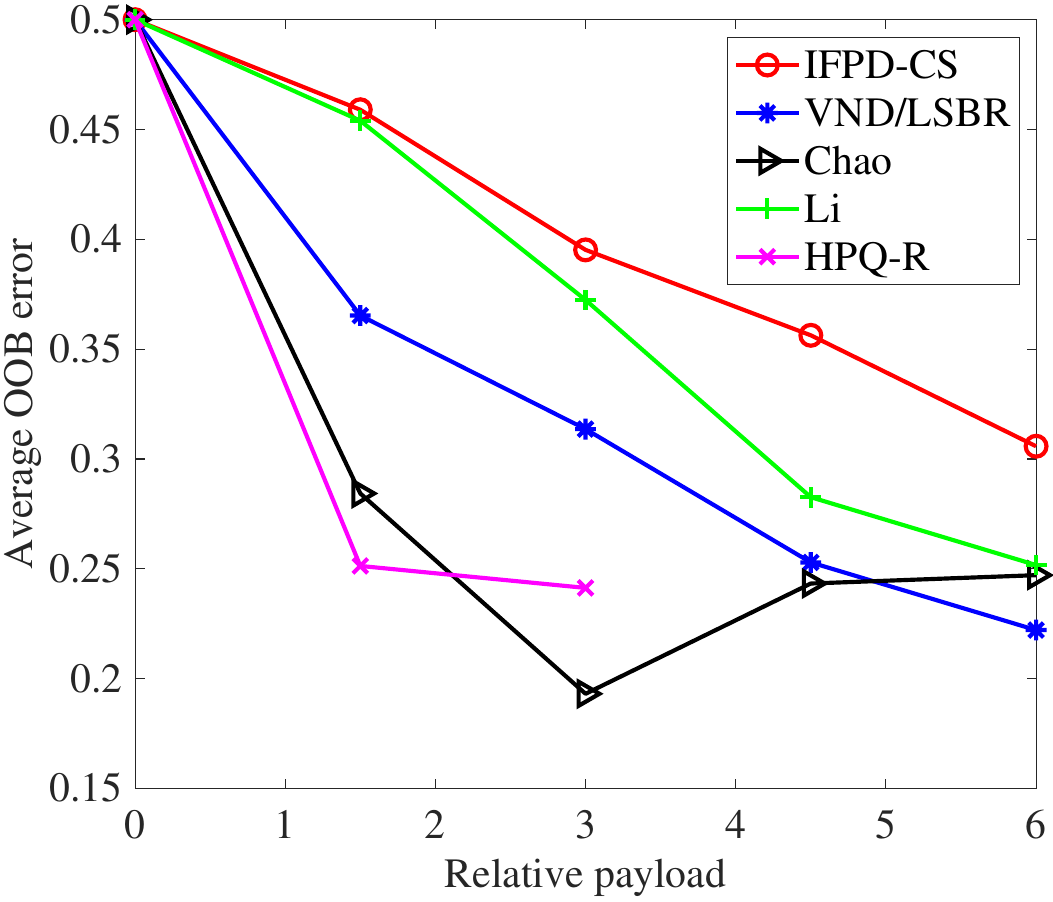}}
	\subfigure[LFS52, PMN dataset]
	{\includegraphics[width=4.2cm]{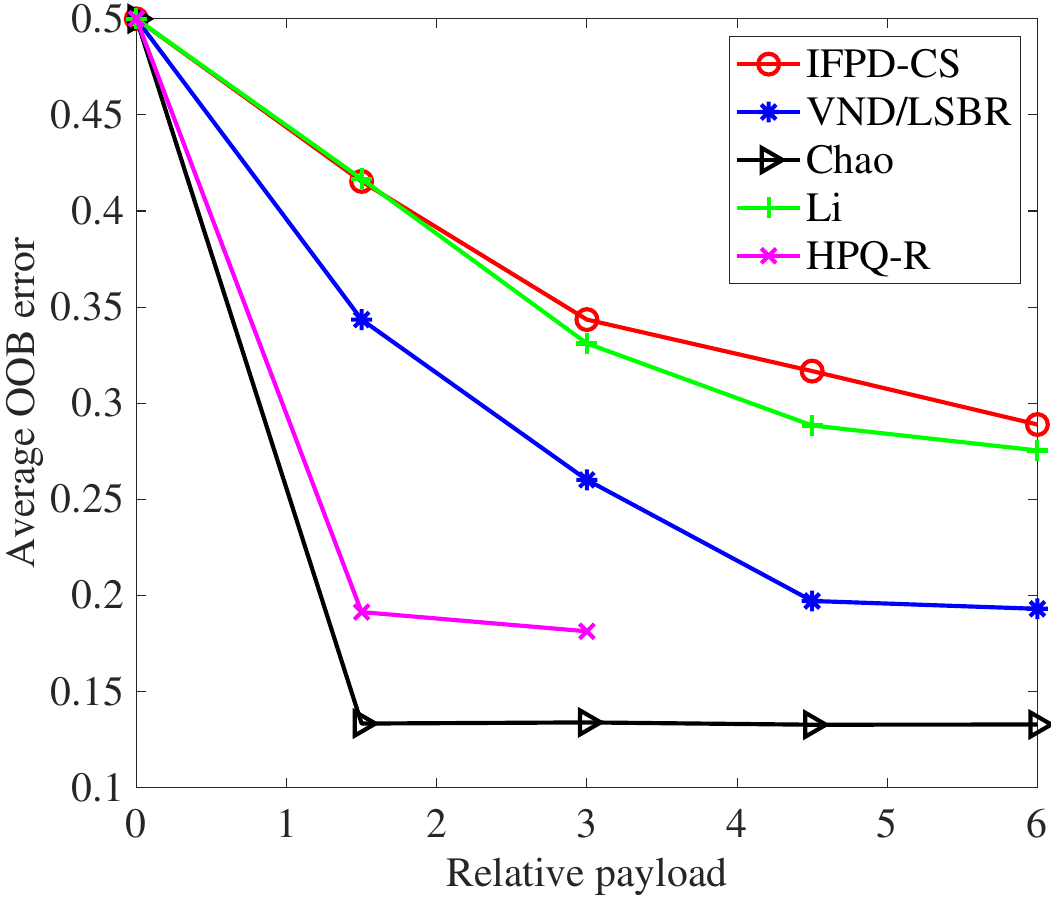}}
	\subfigure[LFS64, PSB dataset] 
	{\includegraphics[width=4.2cm]{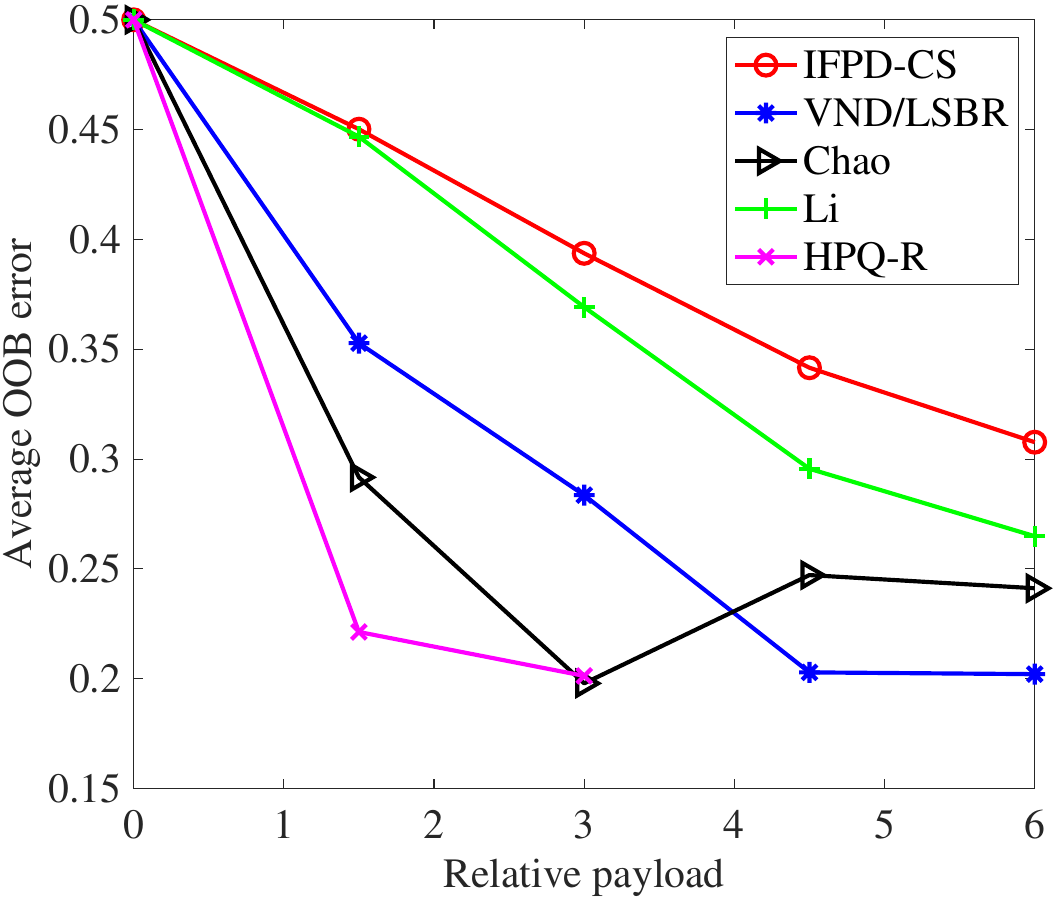}}
	\subfigure[LFS64, PMN dataset]
	{\includegraphics[width=4.2cm]{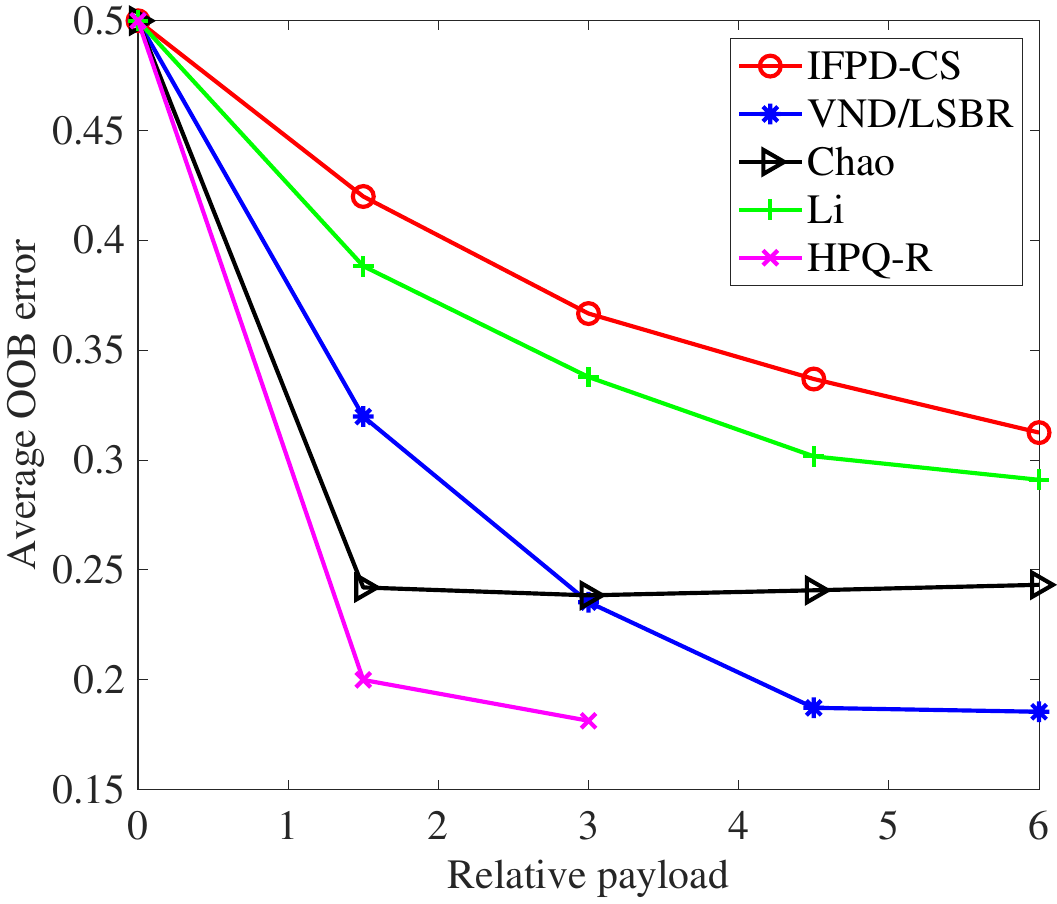}}
	\subfigure[LFS76, PSB dataset] 
	{\includegraphics[width=4.2cm]{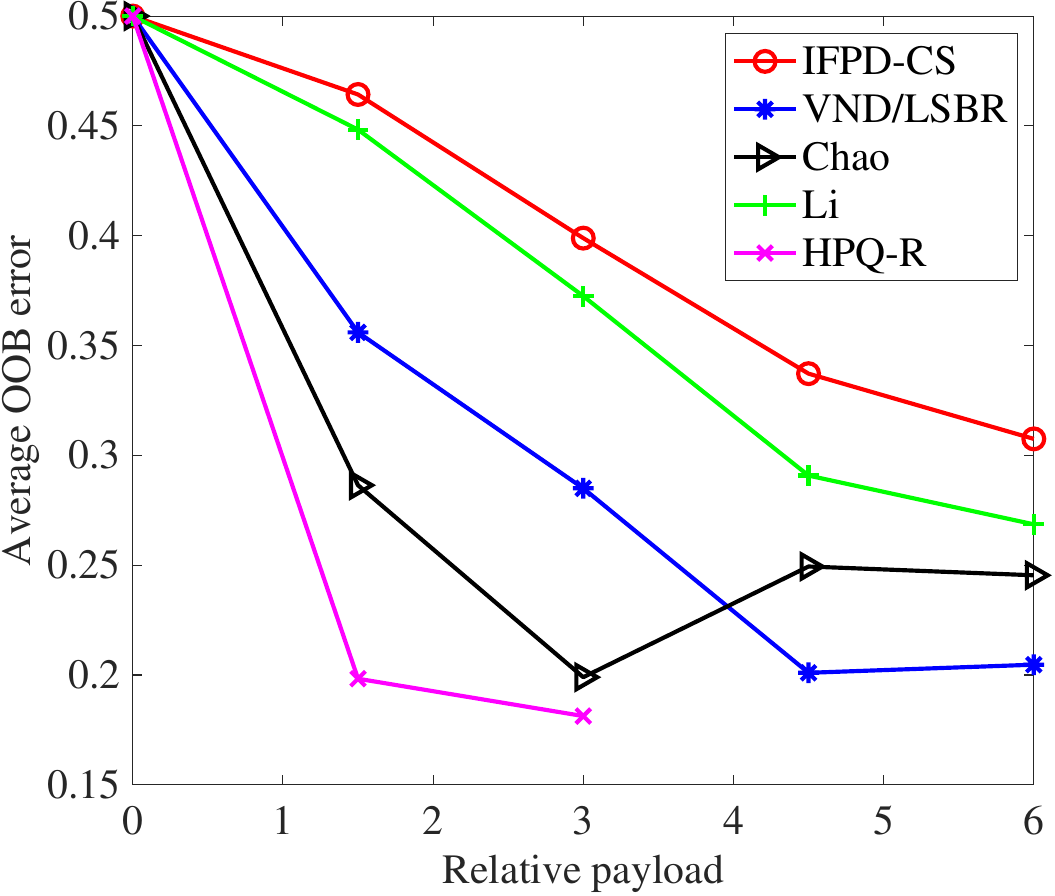}}
	\subfigure[LFS76, PMN dataset]
	{\includegraphics[width=4.2cm]{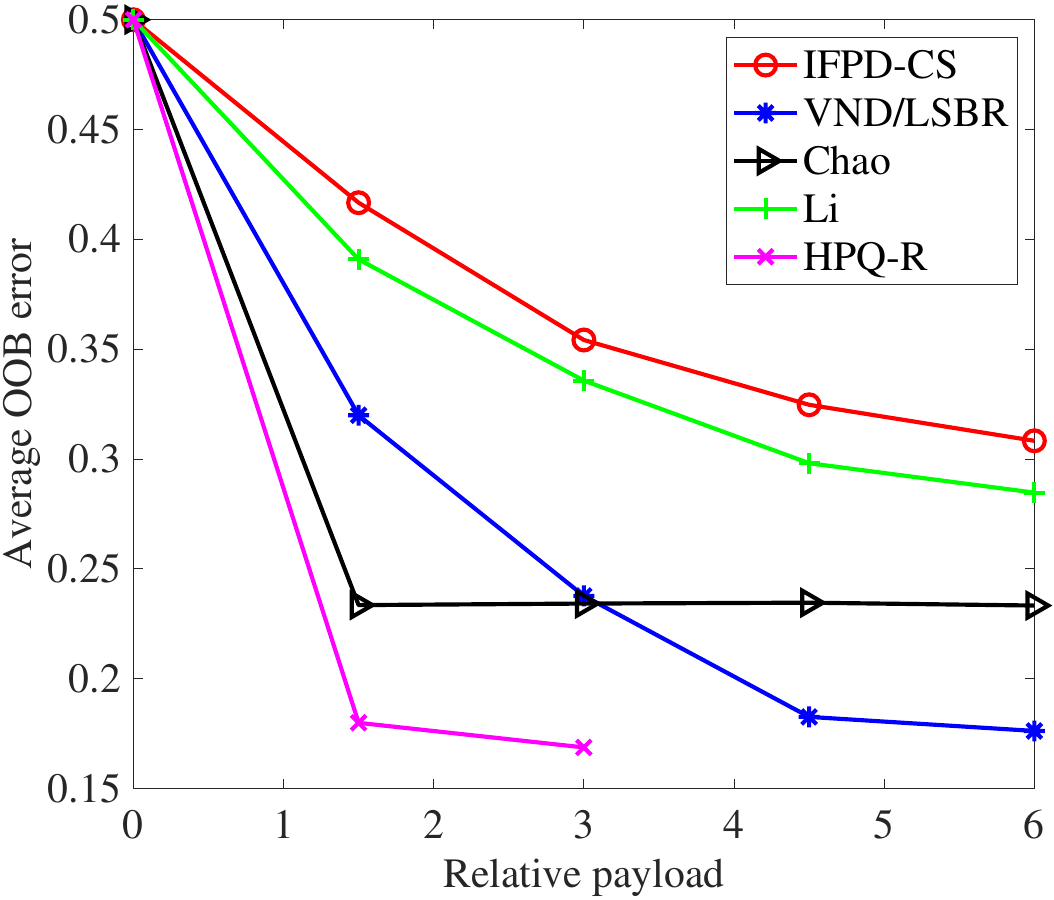}}
	\subfigure[ELFS124, PSB dataset] 
	{\includegraphics[width=4.2cm]{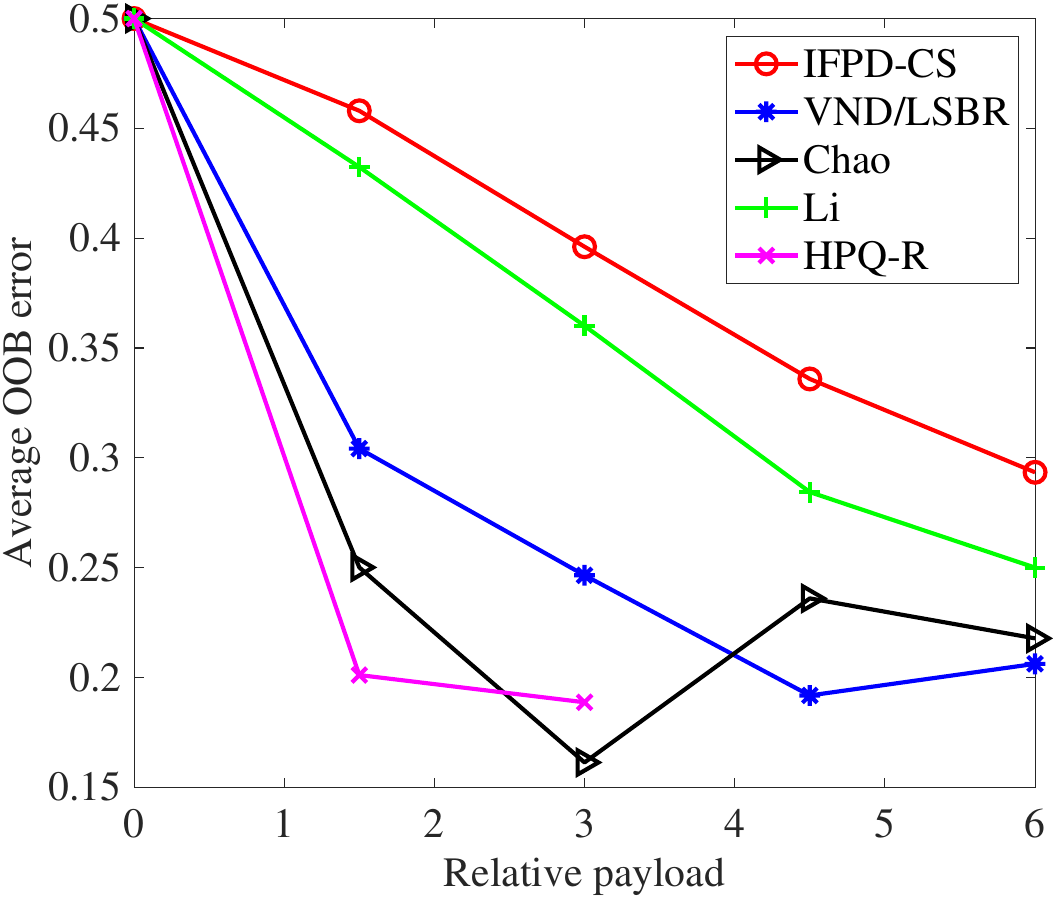}}
	\subfigure[ELFS124, PMN dataset]
	{\includegraphics[width=4.2cm]{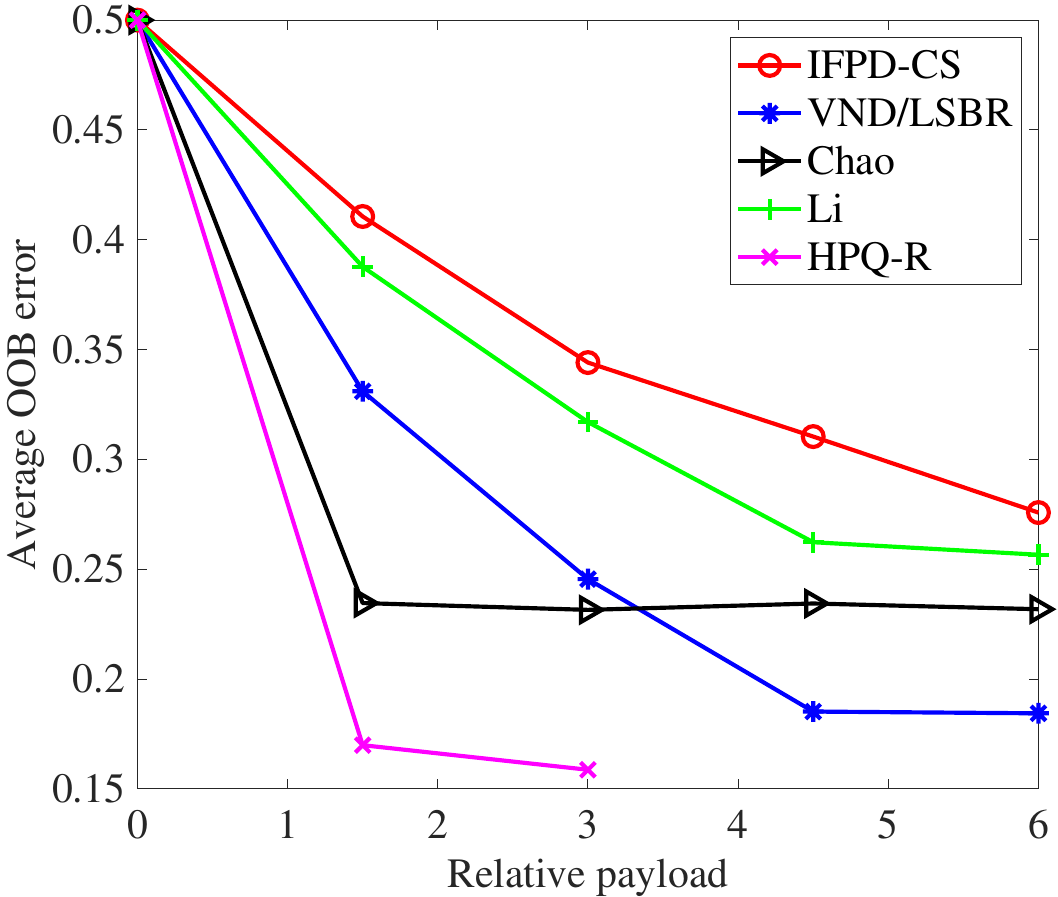}}
	\subfigure[WFS228, PSB dataset] 
	{\includegraphics[width=4.2cm]{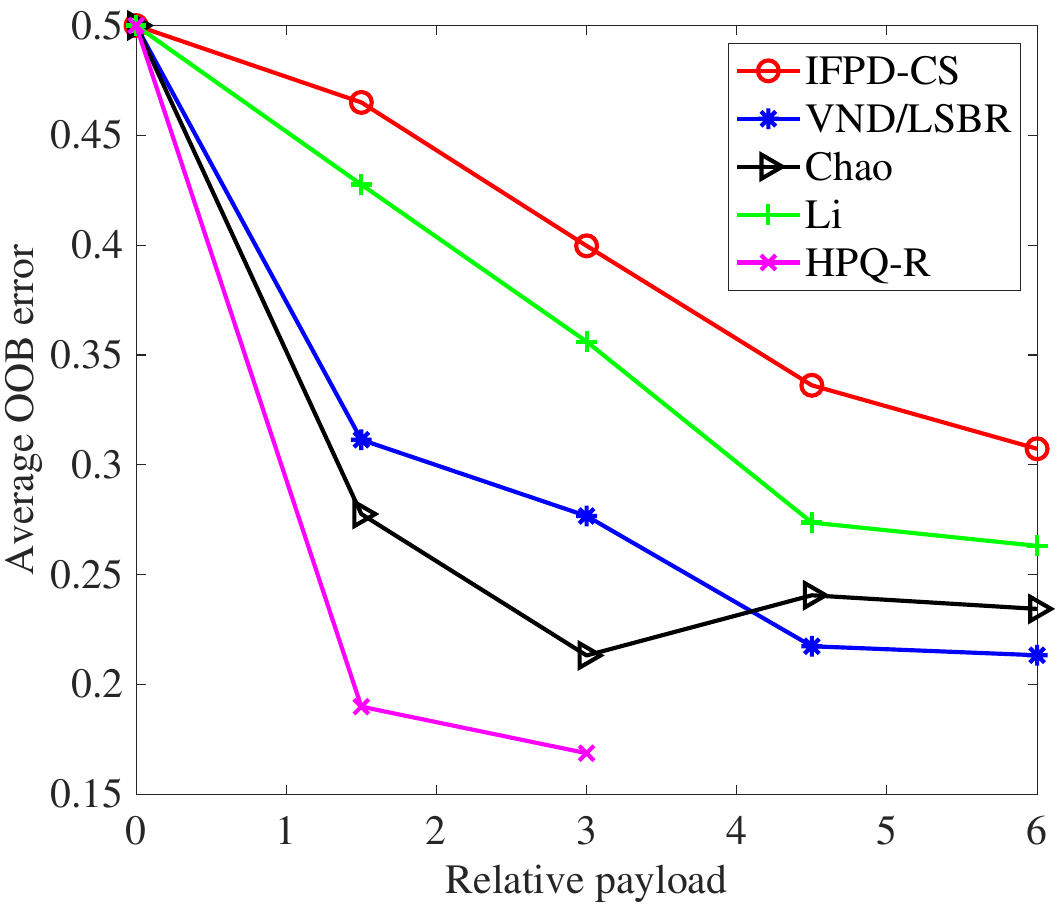}}
    \subfigure[WFS228, PMN dataset] 
	{\includegraphics[width=4.2cm]{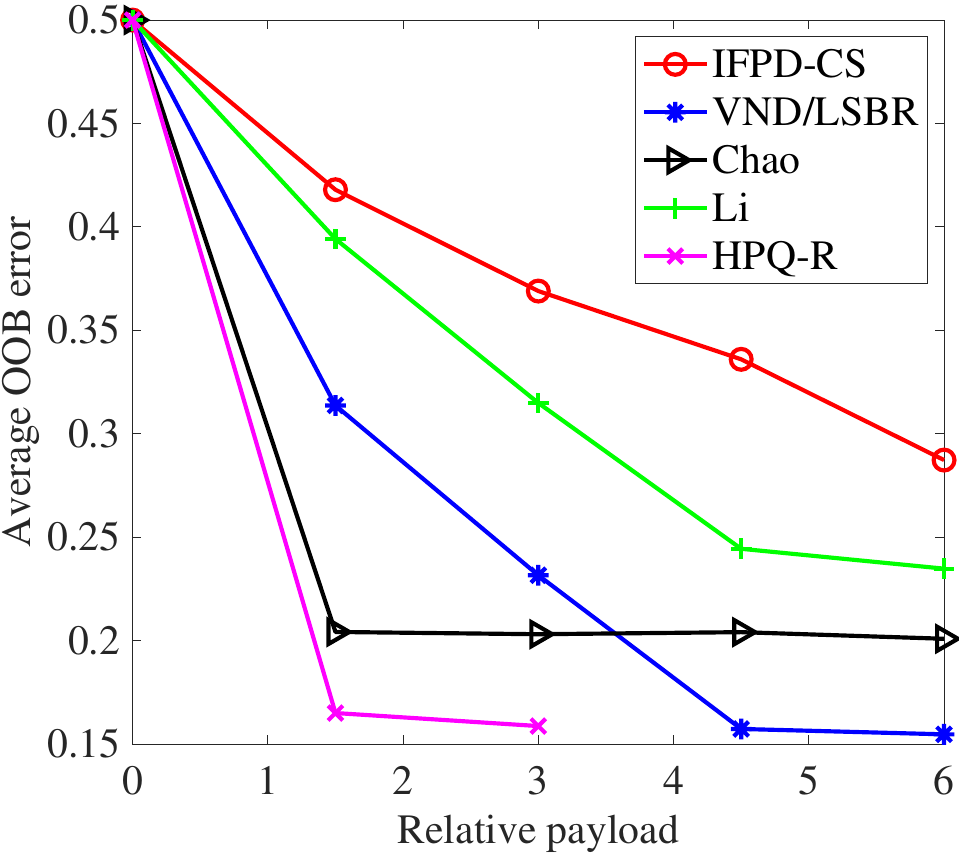}}
	\subfigure[NVT+, PSB dataset]
	{\includegraphics[width=4.2cm]{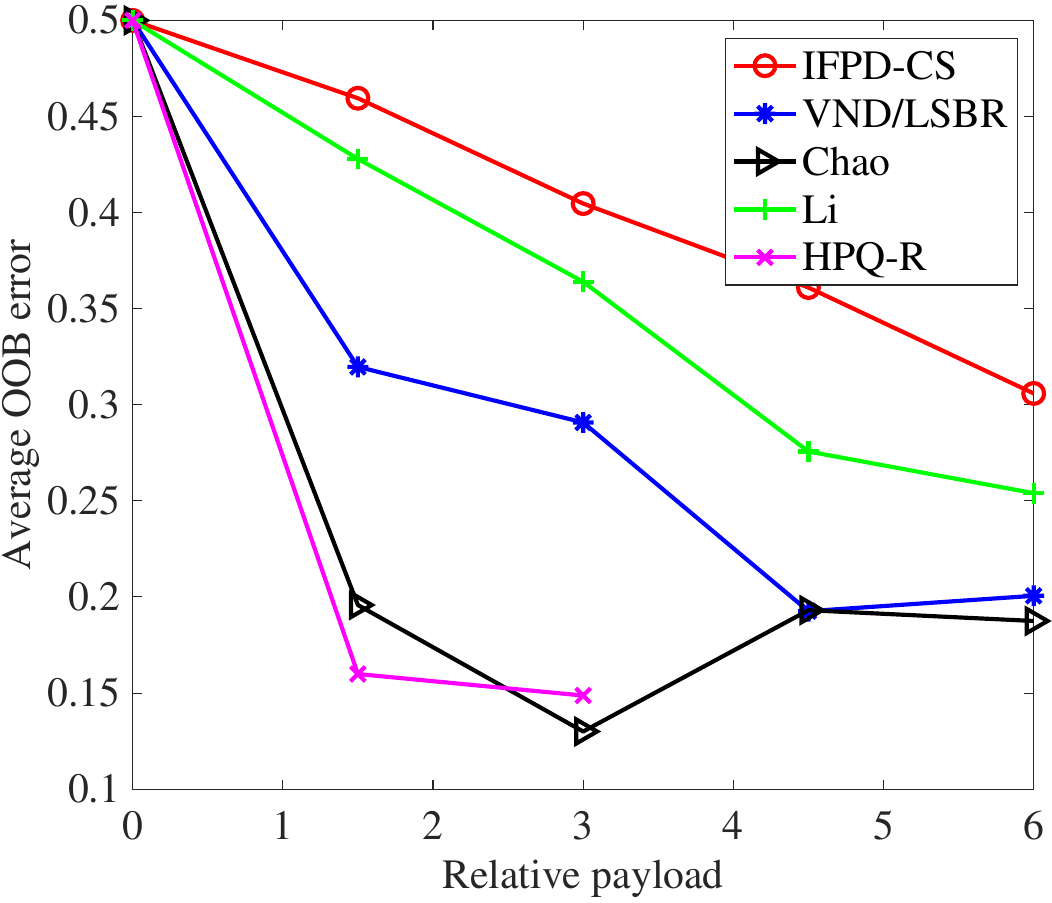}}
	\subfigure[NVT+, PMN dataset] 
	{\includegraphics[width=4.2cm]{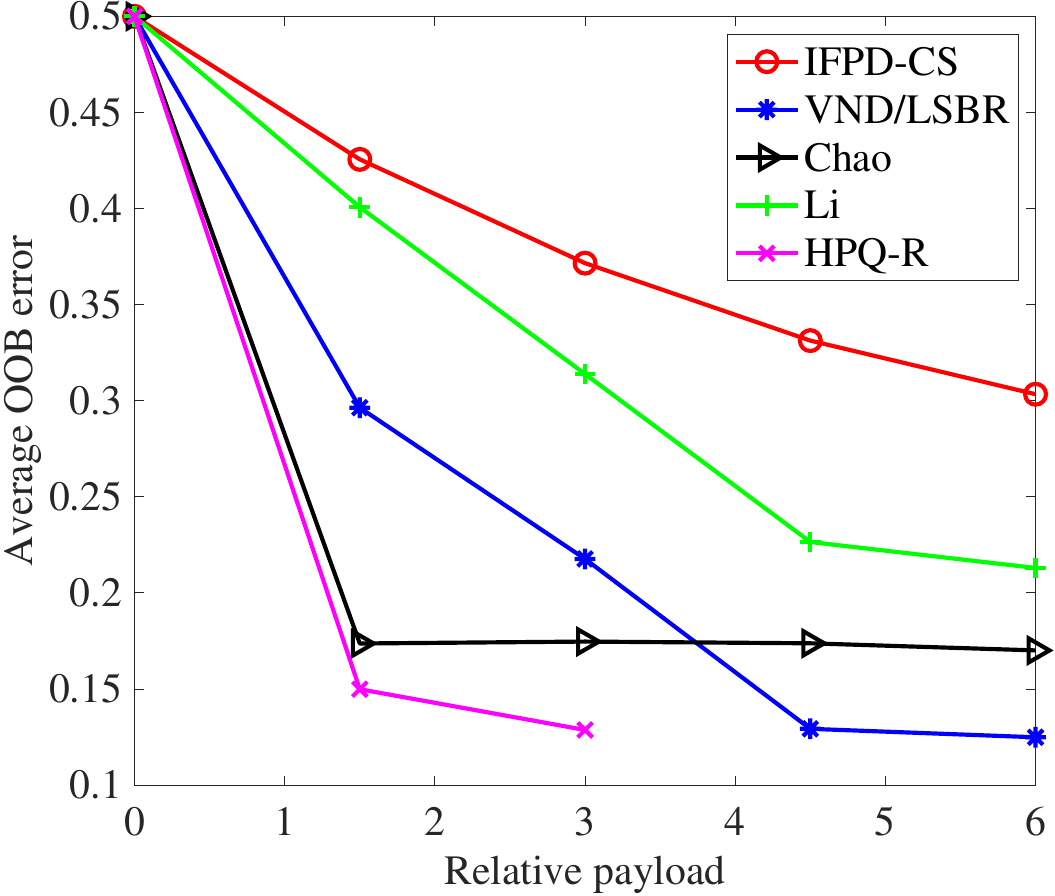}}
	\caption{Average OOB error $\bar{E}_{OOB}$ obtained across different steganalyzers as a function of the relative payload for IFPD-CS, Chao, Li, VND/LSBR, and HPQ-R on the PSB dataset and PMN dataset.} 
	\label{e2}
	\vspace{-1em}
\end{figure*}
\subsubsection{Evaluations of Different Distortion Models}To evaluate the efficacy of the FPD, we opt to compare it with the vertex normal based-distortion (VND) and Gaussian curvature-based distortion (GCD) \cite{VND2018}. In \cite{VND2018}, the two distortion models are defined by
\begin{equation}\label{VNDGCD}
	\begin{aligned}
		&VND(\boldsymbol{v}_i)=\frac{1}{log(||n(\boldsymbol{v}_i)-R(n(\boldsymbol{v}_i))||_2+1)+\sigma}\\
		&GCD(\boldsymbol{v}_i)= \frac{1}{K(\boldsymbol{v}_i)^\beta+\sigma}
	\end{aligned},
\end{equation}
where $n(\boldsymbol{v}_i)$ denotes the vertex normal of $\boldsymbol{v}_i$ and $R(n(\boldsymbol{v}_i))$ is its reference counterpart; $K(\boldsymbol{v}_i)$ is the discrete Gaussian curvature at $\boldsymbol{v}_i$; and $\sigma$ and $\beta$ are constants stabilizing the numerical calculation. Notably, VND and GCD presume that the nonzero elements in $\boldsymbol{I}$ carry equal costs of change. In the following experiments, in addition to IFPD-S1, IFPD-S2, and IFPD-S3, we also evaluate their merged version IFPD-CS. Four relative payloads of 1.5, 3, 4.5, and 6 bpv are considered in the comparison experiments. Specifically, we set $\boldsymbol{I}=\{0,\frac{1}{10^{k_d}}\}$ for $\alpha=1.5$ bpv and adopt the 1-layered STC for message embedding. For $\alpha=3$ bpv, we restrict $\boldsymbol{I}=\{\frac{-1}{10^{k_d}},0,\frac{1}{10^{k_d}},\frac{2}{10^{k_d}}\}$ and perform embedding with 2-layered STC. $\boldsymbol{I}=\{\frac{-3}{10^{k_d}},\frac{-2}{10^{k_d}},\frac{-1}{10^{k_d}},0,\frac{1}{10^{k_d}},\frac{2}{10^{k_d}},\frac{3}{10^{k_d}},\frac{4}{10^{k_d}}\}$ and $\boldsymbol{I}=\{\frac{-7}{10^{k_d}},\cdots,\frac{-1}{10^{k_d}},0,\frac{1}{10^{k_d}},\cdots,\frac{7}{10^{k_d}},\frac{8}{10^{k_d}}\}$ are used for relative payloads of 4.5 and 6 bpv, respectively. Correspondingly, 3-layered STC and 4-layered STC are utilized. We use the STC toolbox accessible on website$\textsuperscript{\ref{f1}}$ and set the height of parity-check submatrix $\widehat{\boldsymbol{\rm{H}}}$ as 10 and the width as 2. The normalization function's form in Eq. (\ref{CF}) can be chosen, and in this work, it is selected as
	$\frac{\rho(\delta_{ij})-\rho_{min}}{\rho_{max}-\rho_{min}}$,
where $\rho_{min}$ and $\rho_{max}$ represent the minimum and maximum values of $\rho(\delta_{ij})$, respectively. Additionally, the scalar $\mu$ in Eq. (\ref{CF}) is set to 1, and the quantization parameter $k_d$ is set to 6. 

In accordance with the aforementioned settings, we conduct experiments on two datasets, namely, PSB and PMN. For the PSB dataset, 280 randomly-selected meshes and their stego counterparts are used as the training set, and the remaining pairs are used for testing purposes. For the PMN dataset, we choose 1280 median-volume meshes as cover meshes, with 1000 cover-stego pairs constituting the training set and the rest reserved for testing. The division of the training and test sets for each dataset is repeated 30 times. To evaluate the distortion models, we employ the 3D steganalyzer NVT+ as the evaluator. Furthermore, we train a separate evaluator for each distortion function and relative payload. The steganography security is assessed by calculating the average $E_{OOB}$ over the 30 test sets, which is denoted by $\bar{E}_{OOB}$. 

Fig. \ref{fig4:a} shows that the steganographic algorithm when using IFPD-S1 outperforms those when using VND and GCD in terms of security, with the difference becoming more significant as the relative payload increases. On the other hand, IFPD-S2, IFPD-S3, and VND perform similarly, while IFPD-CS exhibits remarkably better performance than that of VND and GCD, demonstrating IFPD's effectiveness on the PSB dataset. In Fig. \ref{fig4:b}, IFPD-S1 also improves the steganographic security significantly, but its $\bar{E}_{OOB}$ under each relative payload is lower than that on the PSB dataset. This is because, as described in \cite{NVT2021}, the NVT+ steganalyzer achieves better detection performance on the PMN dataset than that on the PSB dataset. Notably, IFPD-S1, IFPD-S2, and IFPD-S3 exhibit distinct performance on both datasets, with IFPD-S1 exhibiting an absolute advantage in anti-steganalysis, and the other two performing like VND and GCD. This suggests that using IFPD with subfeatures effective for steganalysis does not necessarily improve the algorithm's security significantly, and preserving more features may not be as effective as anticipated. Despite these findings, to avoid overfitting to particular steganalytic features as much as possible \cite{HUGO2010}, we still choose IFPD-CS as the distortion function for the subsequent experiments.
\begin{figure*}[t]
	\centering
	\subfigure[PSB dataset]{\label{fig6:a}\includegraphics[width=0.4\linewidth]{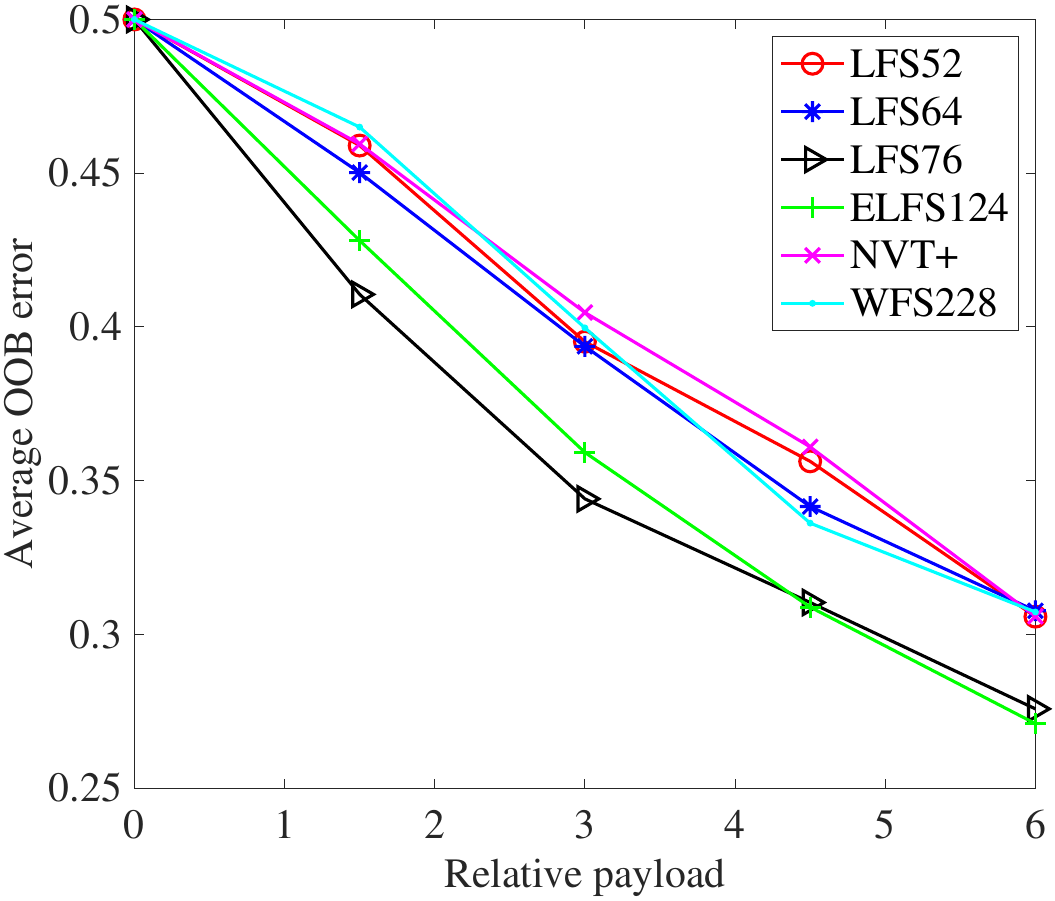}}\hspace{2em}
	\subfigure[PMN dataset]{\label{fig6:b}\includegraphics[width=0.4\linewidth]{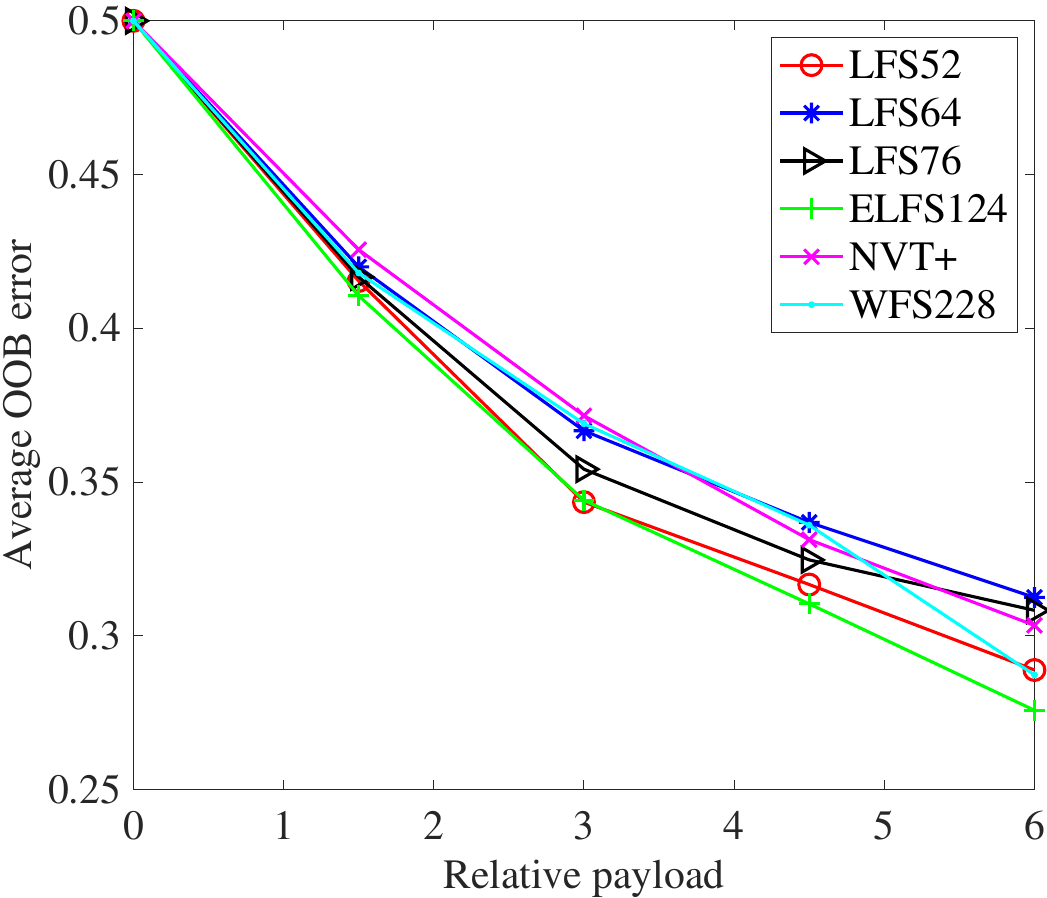}}
	\caption{Average OOB error $\bar{E}_{OOB}$ obtained by IFPD-CS on different steganalyzers as a function of the relative payload on (a) the PSB dataset (b) and PMN dataset.} 
	\label{FPD-se}
\end{figure*}
\subsection{Comparison to State-of-the-Art Algorithms}
To evaluate the anti-steganalysis ability of our algorithm, we compared it with four 3D mesh steganographic algorithms based on geometric modification, namely, VND/LSBR \cite{VND2018}, Chao \cite{Chao2009}, Li \cite{Li2017}, and HPQ-R \cite{LZY2017}.All of these algorithms have high embedding capacities, with VND/LSBR and HPQ-R showing potential resistance to 3D steganalysis. We conducted relevant experiments on the PSB and PMN datasets, with relative payloads of 1.5, 3, 4.5, and 6 bpv. The division of the training and test sets aligns with the description in Section 4.2.2. Since $k_b$ is set to 6, for fair comparison, we configure each algorithm as below. We set Chao's interval number to $10^5$ and the maximum number of embedding layers to 2. For VND/LSBR, we set its initial embedding layer to 14 and reserve the last two layers for LSBR embedding. For Li, we fixed the size of the truncated space as 1 for each vertex. HPQ-R's interval parameter is set to $10^{-5}$, and the number of subintervals is set to $2^3$. It's worth noting that we did not consider HPQ-R at high relative payloads (i.e., 4.5 and 6 bpv) because a higher relative payload requires a larger subinterval number, which may lead to precision loss in vertex coordinates. We utilize the same embedding strategy for each relative payload as described in Section 4.2.2, as well as the STC configuration. To evaluate the steganographic security of each algorithm, we employ six 3D mesh steganalyzers, namely LFS52, LFS64, LFS76, ELFS124, WFS228, and NVT+. The security of each algorithm at each relative payload is assessed by the average OOB error $\bar{E}_{OOB}$ over 30 trials. 
The experimental results are presented in Figs. \ref{e2}. These twelve subfigures demonstrate that IFPD-CS outperforms the other four methods in terms of anti-steganalysis performance on both PSB and PMN. The only exception is Li's method at low relative payloads, where the number of vertices changed by Li is relatively small. However, as the relative payload increases, Li's method shows a progressive decline in $\bar{E}_{OOB}$, and the superiority of IFPD-CS becomes more evident. Another point worth noting is that the $\bar{E}_{OOB}$ curves obtained by our algorithm across various steganalyzers decrease at a relatively slower rate, and even at a relative payload of 6 bpv, IFPD-CS can also achieve an average OOB error of approximately $30\%$ on each steganalyzer. This convinces us that IFPD-CS may maintain good steganography security even at larger relative payloads. 

In Fig. \ref{FPD-se}, we present the average OOB error $\bar{E}_{OOB}$ of the IFPD-CS across different 3D steganalyzers, with relative payloads ranging from 0 to 6 bpv. The steganalysis performances of NVT+ and WFS228 are known to be exceptional compared to those of other 3D steganalyzers, as reported in \cite{NVT2021} and \cite{ZhouS}. However, we consistently observe in Fig. \ref{fig6:a} and Fig. \ref{fig6:b} that they both perform poorly in detecting IFPD-CS on PSB and PMN. This is likely because these two steganalyzers both extract neighborhood-level
representation-guided features for steganalysis, and our algorithm is indeed capable of preserving certain local geometric features. Furthermore, this also highlights a practical issue, namely, that current 3D steganalytic features may not be universal enough.
\subsection{Statistical Significance Test} Considering that the test error distribution of Chao, Li, VND/LSBR, IFPD-CS, and HPQ-R is unknown, we utilize the Wilcoxon rank sum test to evaluate the statistical signiﬁcance of the security improvement achieved by IFPD-CS compared to the other four algorithms under various relative payloads, steganalyzers, and datasets. The hypothesis test is constructed as follows. 
\begin{equation}
	H_0: \mu_1=\mu_2, H_1:\mu_1>\mu_2,
\end{equation}
where $\mu_1$ is the mean values of $n_1$ test OOB errors w.r.t. IFPD-CS; $\mu_2$ is the mean values of $n_2$ test OOB errors of the comparison counterpart. In this case, we set $n_1=n_2=30$. $\mu_1=\mu_2$ indicates that there is no significant difference between them. Let $R$ denote the rank sum of the IFPD-CS test OOB error samples. Since $n_1, n_2 \geq 10$, $R \sim N(\mu_{R},\sigma_{R}^2)$ where $\mu_{R}=\frac{n_1(n_1+n_2+1)}{2}$ and $\sigma^2_{R}=\frac{n_1n_2(n_1+n_2+1)}{12}$. Therefore, we can use
\begin{equation}
	Z=\frac{R-\mu_{R}}{\sigma_{R}}
\end{equation}
as the test statistic. We use the built-in function $ranksum$ of MATLAB to calculate the value of $Z$ and obtain the corresponding p-value. The significance level is set to 0.05. If the p-value is less than the significance level, we reject the null hypothesis $H_0$, indicating that the security improvement of IFPD-CS is statistically significant. Experimental results on the PSB and PMN datasets are presented in Appendix F. In general, our analysis reveals that when compared to Chao, VND/LSBR, Li, and HPD-R, IFPD-CS exhibits a significant improvement in steganography security.
\begin{figure}[!t]
	\centering
	\includegraphics[width=\linewidth]{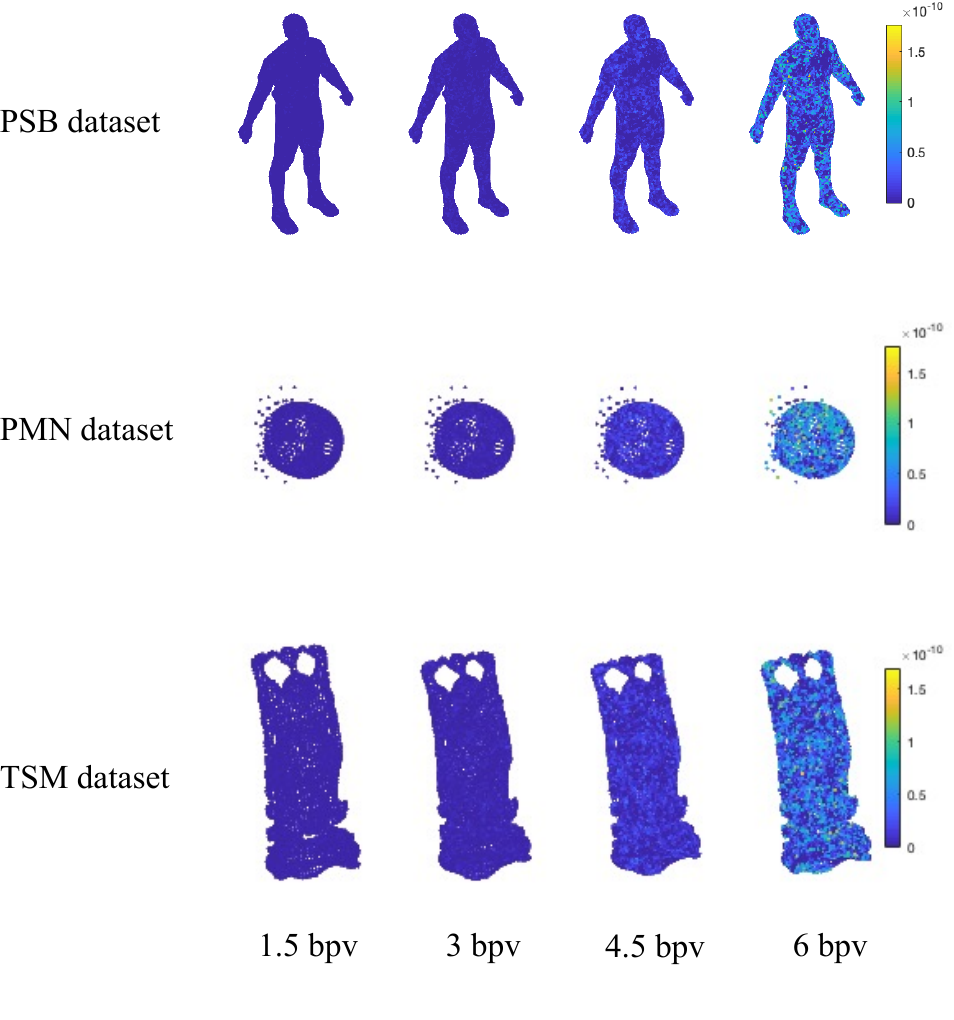}
	\caption{Embedding visualization results generated by our algorithm for meshes ``Human” (PSB), ``Jellyfish” (PMN), and ``Happy Buddha'' (TSM) at different relative payloads.}
	\label{e3a}
\end{figure}
\begin{figure}[!t]
	\centering
	\subfigure[PSB dataset]{\includegraphics[width=\linewidth]{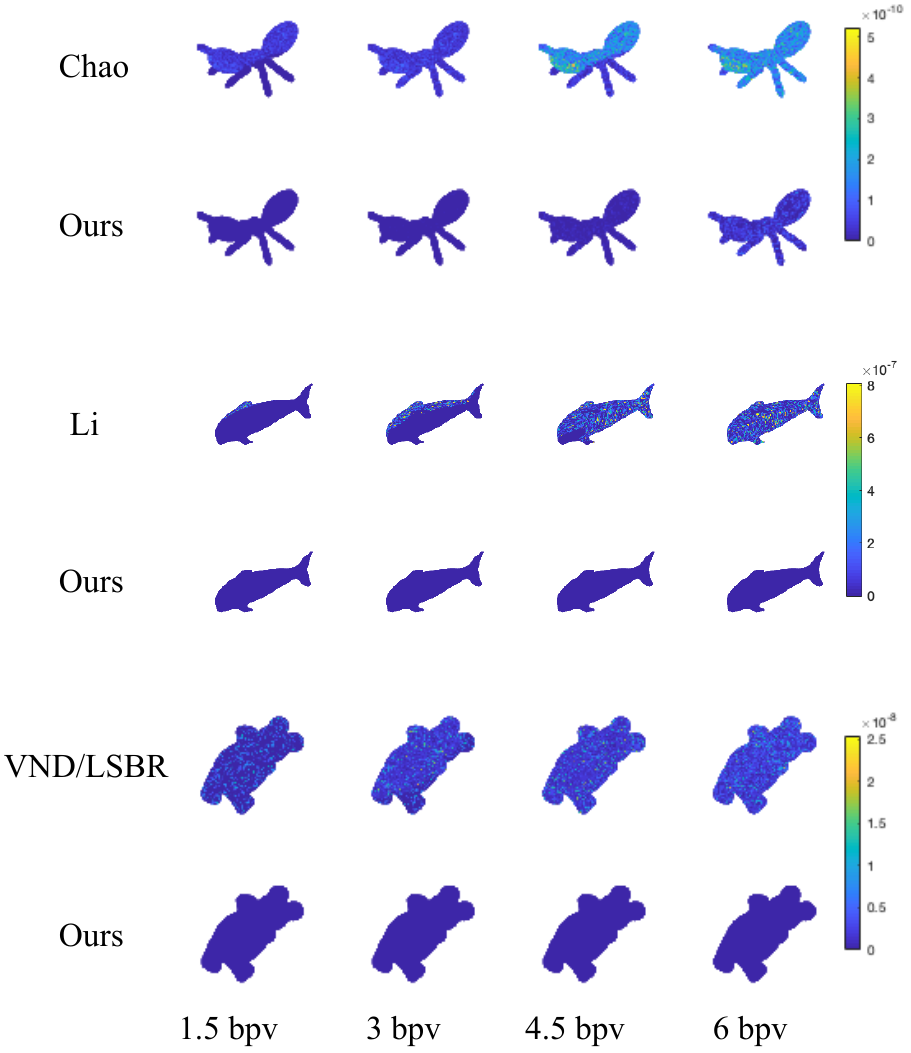}}
	\subfigure[PMN dataset]{\includegraphics[width=\linewidth]{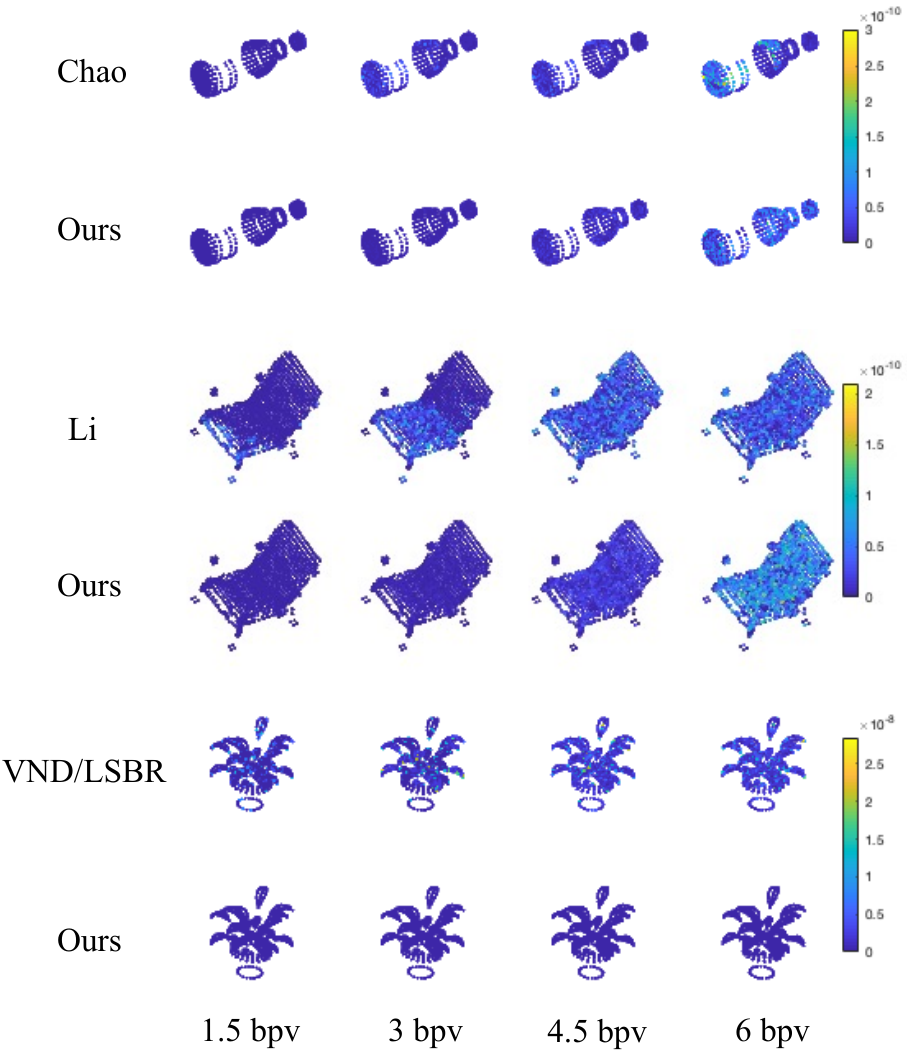}}
	\caption{Visual embedding impact of our algorithm compared to that of Chao, Li, and VND/LSBR for various relative payloads and datasets.}
	\label{e3}
\end{figure}
\subsection{Visualization of Embedding Impact}
We intuitively display several embedding visualization examples from three different datasets in Fig. \ref{e3}. Specifically, we measure the change degree of each vertex by the root-mean-square error and represent it with colors ranging from blue to yellow. We do not show the mesh edges because our algorithm does not change the topology of meshes. To further demonstrate the superiority of our algorithm, we also visualize its message embedding results compared to those of Chao, Li, and VND/LSBR at various relative payloads and on different datasets, as shown in Fig. \ref{e3}. It is evident from the figures in Fig. \ref{e3a} and Fig. \ref{e3} that our algorithm produces more acceptable visual effects of embedding impact on both the PSB and PMN datasets. In particular, stego meshes generated by our method exhibit fewer visual differences than those generated by other algorithms, especially at low relative payloads. This indirectly indicates the capability of our algorithm to preserve certain local and global geometric features, such as sharp features, to a significant extent. Furthermore, it is evident that Chao and Li's message embedding operation falls short in covering all the vertices at times. This observable embedding issue can be traced back to the nonadaptive nature of their algorithms. In sharp contrast to Chao and Li, our algorithm and VND/LSBR algorithms do not display such a phenomenon. Nonetheless, the algorithm VND/LSBR includes the nonadaptive operation LSBR, which makes its embedding impact more conspicuous. Our algorithm assigns each mesh vertex a corresponding cost for vertex modification. In other words, our approach accounts for all vertices and carefully evaluates which vertices can be modified more and which vertices should be modified less during the message embedding process.
\subsection{Analysis of Computational Complexity}
As outlined in Section 3.6, the message embedding procedure comprises two distinct parts: vertex-changing cost calculation and $Q$-layered STC embedding. Without loss of generality, let the time taken to calculate each $\rho(\delta_{ij})$ be $t$. For a mesh with $|\boldsymbol{V}|$ vertices, once $\boldsymbol{I}$ is determined, the time complexity of computing all $\rho(\delta_{ij})$ can be given by $O(3|\boldsymbol{V}||\boldsymbol{I}|t)$. The calculation efficiency of $\rho$, vertex number, and $\boldsymbol{I}$ are the three crucial factors determining the time consumption of the first part. Fortunately, we propose an accelerated cost calculation algorithm that greatly reduces $t$, thereby enhancing the practicality of our algorithm. The second part of the message embedding procedure can be further divided into two parts: BMP calculation and 1-layered STC. Our adoption of acceleration tricks in implementing U$\&$A BMP can make the time taken by BMP calculation negligible. However, the execution time of the STC algorithm is highly dependent on the height of the selected parity-check submatrix. In our work, we select a submatrix with a height of 10, which results in a very short consumption time for STC.
\subsection{Robustness Analysis}In traditional steganography, scholars are more concerned with steganographic security than robustness. However, to move steganography from the laboratory into the real world, a discussion on the robustness of steganography is essential. Zhou \emph{et al.} \cite{ZhouS} enumerated several digital attacks that target 3D mesh steganography, i.e., affine transformation, vertex reordering, smoothing, simplification, and noise addition. Unfortunately, we must admit that our algorithm is not robust enough against these attacks, primarily due to the embedding domain. As we learned from Section 3.6.3, our algorithm's message retrieval procedure is highly sensitive to the modification of vertex coordinates and the change in vertex order, which we overlook when constructing the embedding domain.
\section{Conclusion}
In this paper, we propose a highly adaptive 3D mesh steganographic algorithm based on FPD. Our innovations lie in the construction of the PLS problem, the embedding domain, the distortion function design and improvement, and the BMP calculation for $Q$-layered STC. In Section 4, we also present extensive experiments demonstrating the superiority of our algorithm in resisting steganalysis.

For a long time, research on 3D steganography has been overlooked. However, with the rapid advancement of 3D technology, 3D models are becoming increasingly popular in our daily lives. Hence, we believe that 3D steganography has a bright future. There are still many topics worth exploring in 3D steganography. For instance, the distortion function plays a crucial role in adaptive steganography, and a well-designed distortion function can significantly enhance steganography security. Although the experimental results presented in Fig. \ref{FPD-se} show promising results, we are concerned that the proposed FPD may overfit certain steganalyzers. As a result, designing a more general distortion model is the focus of our future research. Additionally, as highlighted in Section 4.7, our algorithm is susceptible to common attacks due to our constructed embedding domain. Therefore, in our future work, we plan to focus on developing a more robust embedding domain.

%

\ifCLASSOPTIONcaptionsoff
  \newpage
\fi

\appendices
\section{Analysis of NVT+ subfeatures}
Table \ref{tab5} provides the geometric feature relevant to each subfeature of NVT+.

\begin{figure}[!t]
	\centering
	\subfigure[PSB dataset]{\includegraphics[width=8.5cm,height=5cm]{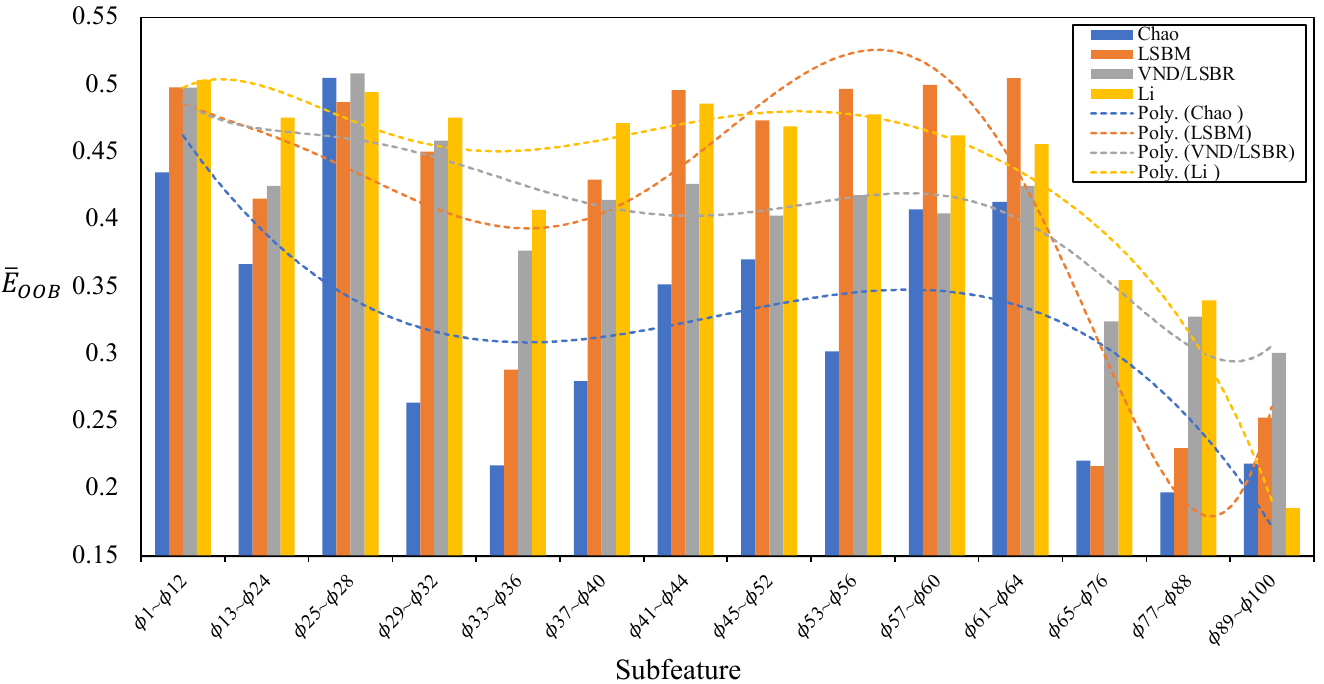}}
	\subfigure[PMN dataset]{\includegraphics[width=8.5cm,height=5cm]{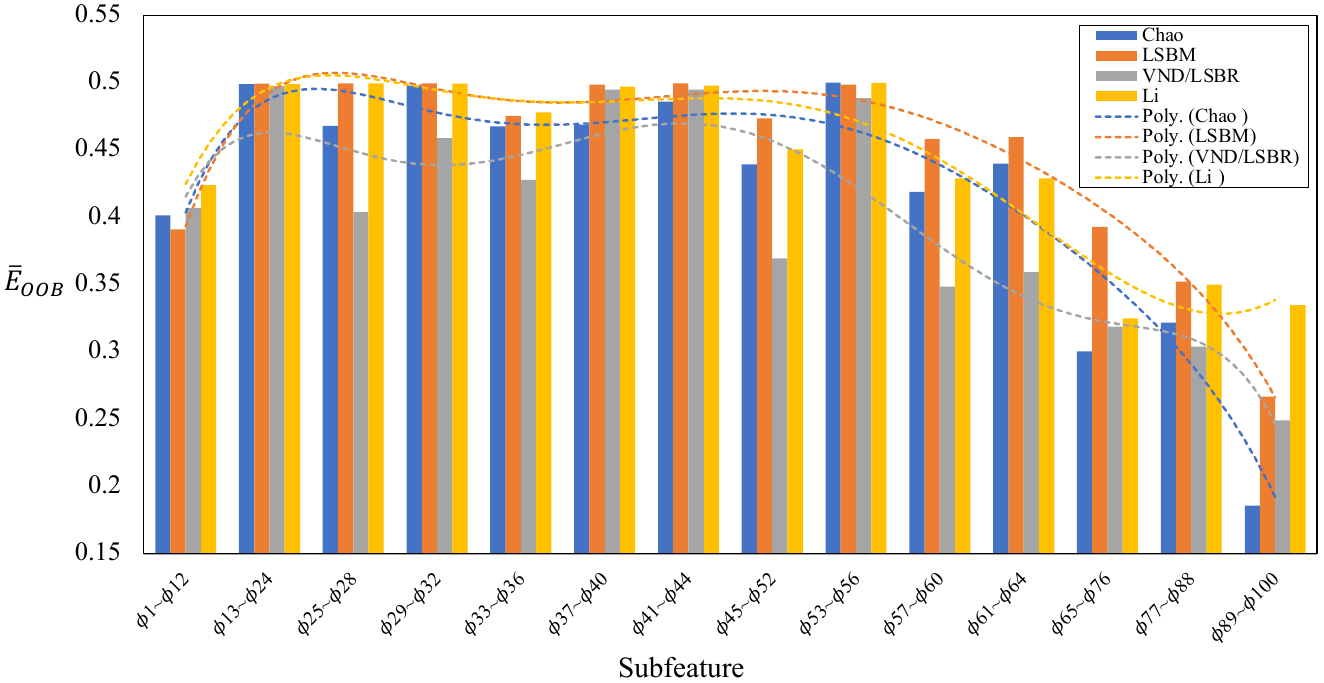}}
	\caption{Average OOB error $\bar{E}_{OOB}$ as a function of the NVT+ subfeature for LSBM, VND/LSBR, Chao, and Li with the relative payload $\alpha$ = 3 bpv on datasets (a) PSB (b) and PMN. Each dotted line in the two figures describes the varying trend of $\bar{E}_{OOB}$ w.r.t. NVT+ subfeatures for different 3D steganography algorithms.}
	\label{SFNVT}
	\vspace{-1em}
\end{figure}

\textbf{Evaluation details}. We evaluate NVT+ features on the Princeton Shape Benchmark (PSB) and Princeton ModelNet (PMN) datasets. For meshes from PSB, we employ LSBM ($\pm 1$), Chao \cite{Chao2009}, VND/LSBR \cite{VND2018}, and Li \cite{Li2017} under relative payload $\alpha=3$ bpv (bits per vertex) to produce their stego counterparts, resulting in four mesh sets, each comprising 380 pairs of cover and stego meshes. For each set,  we randomly select 260 pairs for training and the rest for testing. Such dataset division will be repeated 30 times. For PMN, we randomly select 1280 meshes, of which 1000 are used for training and the rest for testing. The other experimental settings are the same as those on PSB. We adopt the FLD ensemble \cite{FLD2012} for training and testing, and evaluate the test error using the so-called ``out-of-bag" error,  denoted by $E_{OOB}$. The effectiveness of each subfeature under different steganography algorithms is evaluated by computing their average $E_{OOB}$, denoted by $\bar{E}_{OOB}$, over 30 trials, which is reported in Fig. \ref{SFNVT}. In each subfigure, the dotted lines, which are derived from polynomials of order 6, show the varying trends of  $\bar{E}_{OOB}$ w.r.t. different steganography algorithms. 

\textbf{Findings}. It becomes evident in Fig. \ref{SFNVT} (a) that the trend lines of the four steganography algorithms reach their troughs consistently near the subfeature $\phi_{33}\sim\phi_{36}$, and then plummet together near the subfeature $\phi_{65}\sim\phi_{64}$. This indicates that subfeatures $\phi_{33}\sim\phi_{36}$, $\phi_{65}\sim\phi_{76}$, $\phi_{77}\sim\phi_{88}$, and $\phi_{89}\sim\phi_{100}$ are relatively effective for steganalysis on PSB. As for PMN, $\phi_{33}\sim\phi_{36}$ loses its effectiveness, but the precipitous fall occurs again near $\phi_{65}\sim\phi_{76}$, showing that the last three subfeatures are still effective in detecting the four steganography algorithms. In addition, it is a wonder that the subfeature $\phi_{1}\sim\phi_{12}$, which is useless for steganalysis on PSB, bounces back on PMN.
\begin{table}[]	\centering
	\caption{Brief review of the subfeatures of NVT+ in this paper.}
	\label{tab5}
	\begin{tabular}{ll}
		\hline
		Subfeature      & \multicolumn{1}{c}{Relevant geometric feature} \\ \hline
		\multicolumn{1}{l|}{$\phi_{1}\sim\phi_{12}$}   &\multicolumn{1}{c}{\begin{tabular}[c]{@{}l@{}} Vertex coordinates in the Cartesian coordinate system.\end{tabular}}\\
		\multicolumn{1}{l|}{$\phi_{13}\sim\phi_{24}$}   & \multicolumn{1}{c}{\begin{tabular}[c]{@{}l@{}}Vertex coordinates in the Laplacian coordinate system.\end{tabular}}           \\
		\multicolumn{1}{l|}{$\phi_{25}\sim\phi_{28}$} & \multicolumn{1}{c}{\begin{tabular}[c]{@{}l@{}}$l_2$-norm of vertices in the Cartesian coordinate system.\end{tabular}}           \\
		\multicolumn{1}{l|}{$\phi_{29}\sim\phi_{32}$}       &  \multicolumn{1}{c}{\begin{tabular}[c]{@{}l@{}}$l_2$-norm of vertices in the Laplacian coordinate system.\end{tabular}}                                 \\
		\multicolumn{1}{l|}{$\phi_{33}\sim\phi_{36}$}       &  \multicolumn{1}{c}{\begin{tabular}[c]{@{}l@{}}Dihedral angle.\end{tabular}}                                 \\
		\multicolumn{1}{l|}{$\phi_{37}\sim\phi_{40}$}       &   \multicolumn{1}{c}{\begin{tabular}[c]{@{}l@{}}Face normal.\end{tabular}}                                \\
		\multicolumn{1}{l|}{$\phi_{41}\sim\phi_{44}$}       &      \multicolumn{1}{c}{\begin{tabular}[c]{@{}l@{}}Vertex normal.\end{tabular}}                             \\
		\multicolumn{1}{l|}{$\phi_{45}\sim\phi_{52}$}       &    \multicolumn{1}{c}{\begin{tabular}[c]{@{}l@{}}Gaussian curvature and curvature ratio.\end{tabular}}                               \\
		\multicolumn{1}{l|}{$\phi_{53}\sim\phi_{56}$}       &   \multicolumn{1}{c}{\begin{tabular}[c]{@{}l@{}}Edge normal.\end{tabular}}                                \\
		\multicolumn{1}{l|}{$\phi_{57}\sim\phi_{60}$}       &    \multicolumn{1}{c}{\begin{tabular}[c]{@{}l@{}}Mean curvature.\end{tabular}}                               \\
		\multicolumn{1}{l|}{$\phi_{61}\sim\phi_{64}$}       &    \multicolumn{1}{c}{\begin{tabular}[c]{@{}l@{}}Total curvature.\end{tabular}}                               \\
		\multicolumn{1}{l|}{$\phi_{65}\sim\phi_{76}$}       &     \multicolumn{1}{c}{\begin{tabular}[c]{@{}l@{}}NVTs based on $N_1$ (See Fig. \ref{nh}(a)).\end{tabular}}                              \\
		\multicolumn{1}{l|}{$\phi_{77}\sim\phi_{88}$}       &     \multicolumn{1}{c}{\begin{tabular}[c]{@{}l@{}}NVTs based on $N_2$ (See Fig. \ref{nh}(c)).\end{tabular}}                              \\
		\multicolumn{1}{l|}{$\phi_{89}\sim\phi_{100}$}       &     \multicolumn{1}{c}{\begin{tabular}[c]{@{}l@{}}NVTs based on $N_3$ (See Fig. \ref{nh}(e)).\end{tabular}}                              \\
		\hline
	\end{tabular}
\end{table}


%

%
%
%



\begin{figure}[!t]
	\centering
	\includegraphics[width=0.8\linewidth]{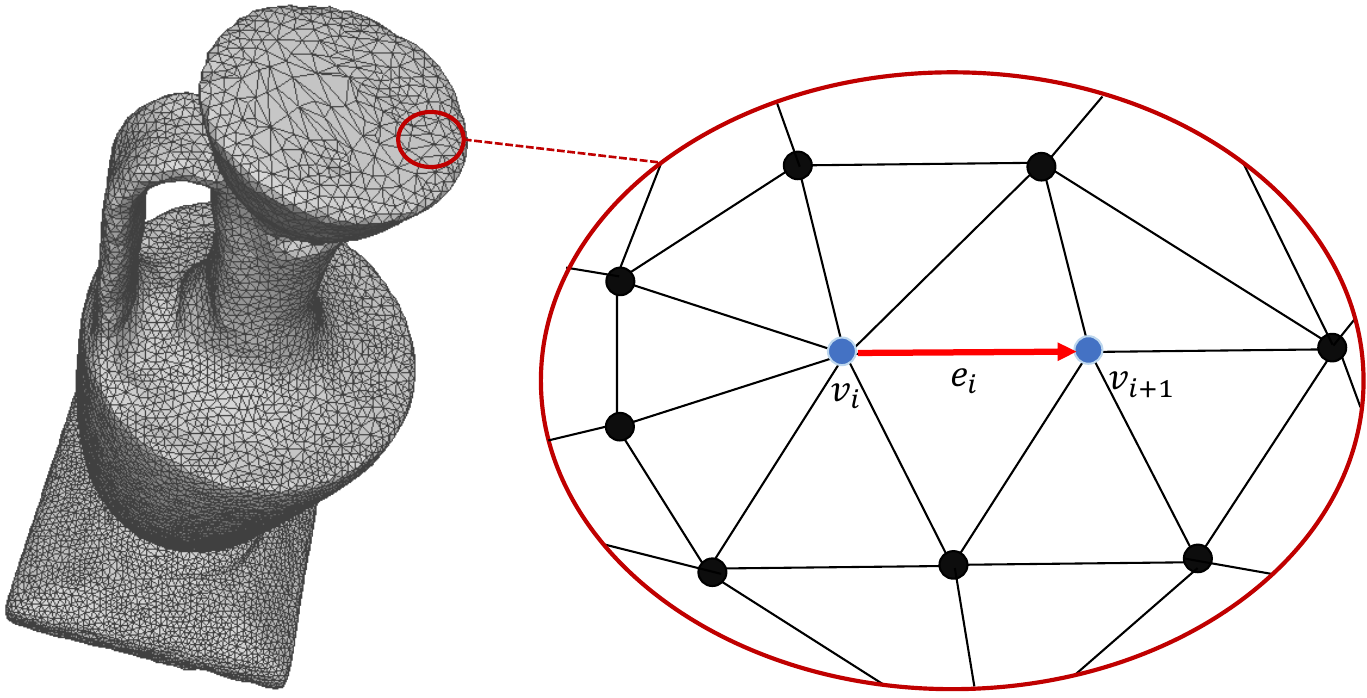}
	\caption{A visual example of a vertex movement that does not change the dihedral angle related features. Note that the zoomed area is a plane.}
	\label{sub1}
\end{figure}
\section{Analysis of subfeatures $\phi_{1} \!\sim \!\phi_{12}$ and $\phi_{33} \!\sim \!\phi_{36}$}
Figure \ref{tab5} illustrates that subfeatures $\phi_{1}\sim\phi_{12}$ and $\phi_{33}\sim\phi_{36}$ share a common issue: they lack generalization for 3D meshes from different datasets. Therefore, including them in the calculation of the cost $\rho(\delta_{ij})$ contradicts the design concept of our FPD. Furthermore, while exhibiting good steganalysis performance on PSB, $\phi_{33} \!\sim \!\phi_{36}$ is still unsuitable for use in calculating $\rho(\delta_{ij})$.
\begin{table*}[!t]
	\begin{center}
		\caption{The (unnormalized) vertex-changing cost of some vertices on a plane in the given mesh when only subfeature $\phi_{33} \!\sim \!\phi_{36}$ is considered in the cost function $\rho$.}
		\label{tab1}
		\resizebox{\textwidth}{25mm}{ 
			\begin{tabular}{ c|  c  c  c  c  c  c  c  c  c  c  c  c  c }
				\hline
				\diagbox{Vertex}{Changes ($/ 10^{k_{IM}}$)} &-6 & -5 & -4 & -3& -2 & -1 & -0& 1 & 2 & 3&4&5&6\\ \hline
				\multicolumn{14}{c}{PSB-346.off} \\
				\hline
				$v_{2601,x}$ & 0& 0 & 0 & 0& 0 & 0& 0& 0& 0 & 0&0&0&0\\ 
				$v_{3455,y}$&3.9969& 3.9887 &  3.9781& 3.9637& 3.9418&3.9003& 0& 3.9003& 3.9418&  3.9637&3.9781& 3.9887&3.9969\\ 
				$v_{3455,z}$& 0 & 0 & 0 & 0& 0 & 0& 0& 0 & 0& 0&0&0&0\\
				\hline 
				\multicolumn{14}{c}{PSB-347.off } \\
				\hline
				$v_{1859,x}$ & 0 & 0 & 0 & 0& 0 & 0& 0& 0 & 0& 0&0&0&0\\ 
				$v_{1859,y}$& 0 & 0 & 0 & 0& 0 & 0& 0& 0 & 0& 0&0&0&0\\ 
				$v_{1859,z}$&5.8171& 5.8027 & 5.7844& 5.7596& 5.7223& 5.6524 & 0& 5.6524 & 5.7223 & 5.7596& 5.7844&5.8027&5.8171\\
				\hline 
				\multicolumn{14}{c}{PSB-153.off} \\
				\hline
				$v_{3362,x}$ & 0.0835 & 0.0869 & 0.0911 & 0.0963& 0.1034 & 0.1147& 0& 0.1147 &  0.1034&0.0963 & 0.0911 &0.0869&0.0835\\
				$v_{3362,y}$& 0 & 0 & 0 & 0& 0 & 0& 0& 0 & 0& 0&0&0&0\\
				$v_{3362,z}$& 0 & 0 & 0 & 0& 0 & 0& 0& 0 & 0& 0&0&0&0\\
				\hline 
				\multicolumn{14}{c}{PSB-104.off} \\
				\hline
				$v_{3119,x}$ & 0 & 0 & 0 & 0& 0 & 0& 0& 0 & 0& 0&0&0&0\\
				$v_{3119,y}$& 0 & 0 & 0 & 0& 0 & 0& 0& 0 & 0& 0&0&0&0\\
				$v_{3119,z}$& 0.3445 & 0.3476 & 0.3512 & 0.3556& 0.3616 & 0.3707& 0& 0.3738 & 0.3616& 0.3556&0.3512&0.3476&0.3445\\
				\hline 
			\end{tabular}
		}
	\end{center}
\end{table*}
A straightforward way to demonstrate this is by providing specific cases where $\phi_{33} \!\sim \!\phi_{36}$ does not work. Specifically, we observe that shifting vertices on a plane in certain directions may not result in any alterations to the features related to dihedral angles. As illustrated in Fig. \ref{sub1}, given two vertices $\boldsymbol{v}_i$ and $\boldsymbol{v}_{i+1}$ on a plane, if we $\boldsymbol{v}_i$ towards $\boldsymbol{v}_{i+1}$ along direction $\boldsymbol{e}_i$, the dihedral angle between every two contiguous faces will never change. This means, for any vertex $\boldsymbol{v}_i$, as long as it is not moved out of the plane, $\rho(\delta_{ij})         \equiv0$ (when only $\phi_{33}\sim\phi_{36}$ is considered in $\rho$). Table \ref{tab1} further confirms this point.  As we know, vertex movement will inevitably bring about changes in mesh features. However, in practice, when given a mesh containing planes, we can always devise a 3D steganography in the way described above to disable our FPD. Consequently, we conclude that subfeature $\phi_{33} \!\sim \!\phi_{36}$ may not be entirely robust.
 
Based on the above analysis, we will not consider subfeatures $\phi_{1} \!\sim \!\phi_{12}$ and $\phi_{33} \!\sim \!\phi_{36}$ in $\rho$.
\section{Calculating $\rho$ with the method given in \cite{NVT2021}}\label{AC} 
The weighting coefficients in Eq. (\ref{nvt}) take the following form:
\begin{equation}
\begin{aligned}
&W_1(\boldsymbol{f})=\frac{A(\boldsymbol{f})}{max(A(N_1(\boldsymbol{v}_i)))}e^{-3||c({\boldsymbol{f}})-\boldsymbol{v}_i||_2},\\
&W_2(\boldsymbol{f})=\frac{A(\boldsymbol{f})}{\sum_{\boldsymbol{f}\in N_2(\boldsymbol{v}_i)}A(\boldsymbol{f})},\\
&W_3(\boldsymbol{f})=\frac{A(\boldsymbol{f})}{\sum_{\boldsymbol{f}\in N_3(\boldsymbol{v}_i)}A(\boldsymbol{f})},\\
\end{aligned}    
\end{equation}
where $A$ is an area function and $c({\boldsymbol{f}})$ denotes the barycenter of the face $\boldsymbol{f}$. 

Let $\lambda_{i1},\lambda_{i2}$, and $\lambda_{i3}$ be the three eigenvalues of the NVT $\boldsymbol{\mathbf{T}}_{i}$\footnote{We simplify the notation $\boldsymbol{{\rm T}}_{ki}$ by dropping the subscript $k$, which is because the subsequent feature extraction method applies to the neighborhood patterns $N_1$, $N_2$, and $N_3$.}, where $i\in\{1,\cdots,N\}$. Eq. ({\ref{nvt}}) states that $N=|\boldsymbol{V}|$ holds for the neighborhood pattern $N_1$, while for the other patterns, $N=|\boldsymbol{F}|$. 

Without loss of generality, let $\lambda_{i1} \geq \lambda_{i2} \geq \lambda_{i3} \geq 0$. 
The eigenvalues of all $\boldsymbol{{\rm T}}_{i}$ can form three sets, i.e., $\boldsymbol{\lambda}_1=\{\lambda_{i1}-\lambda_{i2}\}_{i=1}^{N}$, $\boldsymbol{\lambda}_2=\{\lambda_{i2}-\lambda_{i3}\}_{i=1}^{N}$, and $\boldsymbol{\lambda}_3=\{\lambda_{i3}\}_{i=1}^{N
}$. Given a cover mesh $\boldsymbol{M}$ and its modified counterpart $\boldsymbol{M}'(\delta_{ij})$, \cite{NVT2021} first generates their respective reference counterparts (see Section 2.2 for details), and then it calculates their respective residual features by
\begin{equation}\label{resi}
	\begin{aligned}
		&\boldsymbol{r}_t=NL(abs(\boldsymbol{\lambda}_t-R(\boldsymbol{\lambda}_t))) \\
		&\boldsymbol{r}'_t=NL(abs(\boldsymbol{\lambda}'_t-R(\boldsymbol{\lambda}'_t)))
	\end{aligned} \ \ t\in\{1,2,3\},
\end{equation} 
where $NL$ is a non-linear mapping, $abs$ is an element-wise absolute value function, and $R(\boldsymbol{\lambda})$ denotes the reference mesh feature. Next, the moment-related features w.r.t. $\boldsymbol{r}_t$ and $\boldsymbol{r}'_t$ are calculated by
\begin{equation}
    \begin{aligned}
    & \boldsymbol{mo}_t = \{mean(\boldsymbol{r}_t),var(\boldsymbol{r}_t),skewness(\boldsymbol{r}_t),kurtosis(\boldsymbol{r}_t)\},\\
    & \boldsymbol{mo}'_t = \{mean(\boldsymbol{r}'_t),var(\boldsymbol{r}'_t),skewness(\boldsymbol{r}'_t),kurtosis(\boldsymbol{r}'_t)\},
    \end{aligned}
\end{equation}

Give a neighborhood pattern $N_k$, $S_k(\boldsymbol{M})=\{\boldsymbol{mo}_1,\boldsymbol{mo}_2,\boldsymbol{mo}_3\}$ and $S_k(\boldsymbol{M}'(\delta_{ij}))=\{\boldsymbol{mo}'_1,\boldsymbol{mo}'_2,\boldsymbol{mo}'_3\}$. By bringing them into Eq. (\ref{CF}), we can obtain the complete form of the cost function $\rho$. By repeating the above steps, we can calculate $\rho(\delta_{ij})$ for all vertices and $\delta_{ij}\in \boldsymbol{I}$. 
\section{Further Explanation of Remark 1} 
As $\boldsymbol{M}$ is watertight, we can get $|ID_1(\boldsymbol{v}_i)|=|N_1(\boldsymbol{v}_i)|+1$, where $N_1(\boldsymbol{v}_i)$ is the set of 1-ring neighboring faces w.r.t. $\boldsymbol{v}_i$. Note that $|N_1(\boldsymbol{v}_i)|$ not only represents the number of 1-ring neighboring faces of $\boldsymbol{v}_i$ but also the number of 1-ring neighboring vertices (see Fig. \ref{nh}(a)). In fact, $N_1(\boldsymbol{v}_i)$ is just a tiny patch of $\boldsymbol{M}$. Thus
we can get $|N_1(\boldsymbol{v}_i)|\ll|\boldsymbol{V}| \Leftrightarrow |ID_1(\boldsymbol{v}_i)|\ll|\boldsymbol{V}|+1$, so $|ID_1(\boldsymbol{v}_i)|\ll|\boldsymbol{V}|$. For $ID_2(\boldsymbol{v}_i)$, we have $|N_1(\boldsymbol{v}_i)|\leq|ID_2(\boldsymbol{v}_i)|\leq2|N_1(\boldsymbol{v}_i)|$. As we have known $|N_1(\boldsymbol{v}_i)|\ll|\boldsymbol{F}|$, so $|ID_2(\boldsymbol{v}_i)|\ll|\boldsymbol{F}|$. As for $ID_3(\boldsymbol{v}_i)$, we have 
$|N_1(\boldsymbol{v}_i)|\leq|ID_3(\boldsymbol{v}_i)|\leq|ID_2(\boldsymbol{v}_i)|+NF$, where $NF$ actually denotes the number of uncolored faces within the red dash lines in Fig. \ref{nh}(d). Based on the assumption that the neighbor number of each vertex in $\boldsymbol{M}$ is about the same, we have $NF\ll|\boldsymbol{F}|$, and finally get $|ID_3(\boldsymbol{v}_i)|\ll|\boldsymbol{F}|$.
\begin{algorithm}[!t]
	\caption{Accelerated Vertex-changing Cost Calculation} 
	\hspace*{0.02in} {\bf Input:} 
	Mesh $\boldsymbol{M}=\{\boldsymbol{V},\boldsymbol{E},\boldsymbol{F}\}$, neighborhood pattern $\boldsymbol{N}_k$, perturbation set $\boldsymbol{I}$.\\
	\hspace*{0.02in} {\bf Output:} 
	Vertex-changing cost $\{\rho(\delta_{ij})|1\leq i\leq|\boldsymbol{V}|,j\in\{x,y,z\},\delta_{ij}\in \boldsymbol{I}\}$.
	\begin{algorithmic}[1]
   \State Obtain face normal set $\boldsymbol{fn}$, face barycenter set $\boldsymbol{fb}$ (needed only for $N_1$), face area set $\boldsymbol{fa}$ of $\boldsymbol{M}$;
		\State Calculate eigenvalue related features $\boldsymbol{\lambda}_t$ of $\boldsymbol{M}, t\in\{x,y,z\}$;
  \State Calculate eigenvalue related features $R(\boldsymbol{\lambda}_t)$ of the reference of $\boldsymbol{M}, t\in\{x,y,z\}$;
       \State Calculate $S_k(\boldsymbol{M})$ according to Appendix C;
		\For{$i=1$ to $|\boldsymbol{V}|$}
                \For{$j$ in $\{x,y,z\}$}
        \For{$\delta_{ij}$ in $\boldsymbol{I}$}
                 \State Recalculate the normals, barycenters (needed only for $N_1$), and areas of faces in $N_1(\boldsymbol{v}_i)$, obtaining three subsets $\tilde{\boldsymbol{fn}}$, $\tilde{\boldsymbol{fb}}$, and $\tilde{\boldsymbol{fa}}$.
                 \State Replace the corresponding items in $\boldsymbol{fa}$, $\boldsymbol{fb}$, and $\boldsymbol{fc}$ with those in $\tilde{\boldsymbol{fn}}$, $\tilde{\boldsymbol{fb}}$ (needed only for $N_1$), and $\tilde{\boldsymbol{fa}}$, respectively.
                \State Obtain the NVTs that correspond to the elements in $ID_k(\boldsymbol{v}_i)$ and calculate their eigenvalues, resulting in three features $\tilde{\boldsymbol{\lambda}}_1$, $\tilde{\boldsymbol{\lambda}}_2$, and $\tilde{\boldsymbol{\lambda}}_3$;
                \State Replace the corresponding items in $\boldsymbol{\lambda}_t$ with $\tilde{\boldsymbol{\lambda}}_t$ to form $\boldsymbol{\lambda}'_t$, where $t \in \{1,2,3\}$;
                \State Calculate $\boldsymbol{r}'_t=\boldsymbol{\lambda}'_t-R(\boldsymbol{\lambda}_t), t\in\{1,2,3\}$;
                \State Calculate $S_k(\boldsymbol{M}'(\delta_{ij}))$ with $\boldsymbol{r}'_t$ according to Appendix C;
                \State $\rho(\delta_{ij})=\mu N_{or}(||S_k(\boldsymbol{M})-S_k(\boldsymbol{M}'(\delta_{ij}))||_1)$
                \EndFor
                \EndFor
            \EndFor
		\State \Return $\{\rho(\delta_{ij})|1\leq i\leq|\boldsymbol{V}|,j\in\{x,y,z\},\delta_{ij}\in \boldsymbol{I}\}$.
	\end{algorithmic}
 \label{acc}
\end{algorithm}
\section{Acceleration Algorithm}
The pseudocode of accelerating the calculation of $\rho(\delta_{ij})$ for each vertex in $\boldsymbol{M}$ and $\delta_{ij}\in \boldsymbol{I}$ is shown in Algorithm \ref{acc}. Note that the acceleration paradigm shown in Algorithm \ref{acc} is applicable to any of $\phi_{65}\sim\phi_{76}$, $\phi_{77}\sim\phi_{88}$, and $\phi_{89}\sim \phi_{100}$, so we only provide the acceleration algorithm with only one subfeature considered in $\rho$. Let us delve deeper into this algorithm. The main factors that enable Algorithm \ref{acc} to accelerate the calculation of vertex-changing cost is Line 10. This is because it solely focuses on the calculation of eigenvalues of a tiny portion of NVTs determined by $ID_k(\boldsymbol{v}_i)$, as expressed in Remark \ref{p2}. Furthermore, we approximate $R(\boldsymbol{\lambda}'_t)$ with $R(\boldsymbol{\lambda}_t)$ in Line 12 because we find that the reference of $\boldsymbol{M}$ is extremely similar to that of $\boldsymbol{M}(\delta_{ij})$. In practical code, we also implement some acceleration tricks for recalculating the weighting coefficients $W_k(\boldsymbol{f})$ in Lines 8 and 9. For more implementation details, please refer to our open-source code\footnote{https://github.com/zjhJOJO/3D-steganography-based-on-FPD}.
\section{Experimental Results of Statistical Significance Test}
Experimental results on the PSB and PMN datasets are presented in Table \ref{tab3} and Table \ref{tab4}, respectively. P-values greater than the significance level are shown in bold font. As shown in these two tables, our algorithm obtain a significant improvement in security when compared to Chao, VND/LSBR, and HPD-R, IFPD-CS provides. However, in the case of Li at low relative payloads, the improvement is not always significant, which is consistent with the experimental results presented in Fig. \ref{e2}.
\begin{table*}[t]
	\centering
	\caption{The p-values of IFPD-CS compared to those of Chao, Li, VND/LSBR, and HPQ-R on the PSB dataset for various relative payloads and steganalyzers.}
	\label{tab3}
	\begin{tabular}{ cl|cccc }
		\hline
		&&Chao&Li&VND/LSBR&HPQ-R\\
		\hline
		\multirow{4}{*}{LFS52} 
		& $\alpha=1.5$ &$8.4556\times10^{-18}$&\textbf{0.9998}&$8.4556\times10^{-18}$&$8.4556\times10^{-18}$\\
		& $\alpha=3$ & $8.4556\times10^{-18}$&$2.1724\times10^{-10}$&$8.4556\times10^{-18}$&$8.4556\times10^{-18}$\\
		& $\alpha=4.5$ & $8.4556\times10^{-18}$&$5.5283\times10^{-14}$&$8.4556\times10^{-18}$&- \\
		& $\alpha=6$ & $8.4556\times10^{-18}$&$3.4600\times10^{-12}$&$8.4556\times10^{-18}$&- \\
		\hline
		\multirow{4}{*}{LFS64} 
		& $\alpha=1.5$&$8.4556\times10^{-18}$&\textbf{0.0766}&$8.4556\times10^{-18}$&$8.4556\times10^{-18}$\\
		&$\alpha=3$&$8.4556\times10^{-18}$&$1.4271\times10^{-8}$&$8.4556\times10^{-18}$&$8.4556\times10^{-18}$\\
		& $\alpha=4.5$ & $8.4556\times10^{-18}$&$9.7805\times10^{-12}$&$8.4556\times10^{-18}$&- \\
		& $\alpha=6$ & $8.4556\times10^{-18}$&$1.2880\times10^{-8}$&$8.4556\times10^{-18}$&- \\
		\hline
		\multirow{4}{*}{LFS76} 
		& $\alpha=1.5$&$8.4556\times10^{-18}$&\textbf{0.7130}&$8.4556\times10^{-18}$&$8.4556\times10^{-18}$\\
		&$\alpha=3$&$8.4556\times10^{-18}$&$3.0429\times10^{-5}$&$8.4556\times10^{-18}$&$8.4556\times10^{-18}$\\
		& $\alpha=4.5$ & $8.4556\times10^{-18}$&$4.1049\times10^{-13}$&$8.4556\times10^{-18}$&- \\
		& $\alpha=6$ & $8.4556\times10^{-18}$&$5.7485\times10^{-11}$&$8.4556\times10^{-18}$&- \\
		\hline
		\multirow{4}{*}{ELFS124} 
		& $\alpha=1.5$&$8.4556\times10^{-18}$&$2.0954\times10^{-6}$&$8.4556\times10^{-18}$&$8.4556\times10^{-18}$\\
		&$\alpha=3$&$8.4556\times10^{-18}$&0.0036&$8.4556\times10^{-18}$&$8.4556\times10^{-18}$\\
		& $\alpha=4.5$ & $8.4556\times10^{-18}$&$1.6684\times10^{-9}$&$8.4556\times10^{-18}$&- \\
		& $\alpha=6$ & $8.4556\times10^{-18}$&$6.7545\times10^{-6}$&$8.4556\times10^{-18}$&- \\
		\hline
		\multirow{4}{*}{NVT+} 
		& $\alpha=1.5$&$8.4556\times10^{-18}$&$2.1823\times10^{-8}$&$8.4556\times10^{-18}$&$8.4556\times10^{-18}$\\
		&$\alpha=3$&$8.4556\times10^{-18}$&$1.8789\times10^{-11}$&$8.4556\times10^{-18}$&$8.4556\times10^{-18}$\\
		& $\alpha=4.5$ & $8.4556\times10^{-18}$&$2.5367\times10^{-17}$&$8.4556\times10^{-18}$&- \\
		& $\alpha=6$ & $8.4556\times10^{-18}$&$1.0705\times10^{-14}$&$8.4556\times10^{-18}$&- \\
		\hline
		\multirow{4}{*}{WFS228} 
		& $\alpha=1.5$&$8.4556\times10^{-18}$&$1.7130\times10^{-13}$&$8.4556\times10^{-18}$&$8.4556\times10^{-18}$\\
		&$\alpha=3$&$8.4556\times10^{-18}$&$3.8482\times10^{-10}$&$8.4556\times10^{-18}$&$8.4556\times10^{-18}$\\
		& $\alpha=4.5$ & $8.4556\times10^{-18}$&$1.5220\times10^{-16}$&$8.4556\times10^{-18}$&- \\
		& $\alpha=6$ & $8.4556\times10^{-18}$&$1.1112\times10^{-13}$&$8.4556\times10^{-18}$&- \\
		\hline
	\end{tabular}
\end{table*}
\begin{table*}[t]
	\centering
	\caption{P-values of IFPD-CS compared to Chao, Li, VND/LSBR, and HPQ-R on the PMN dataset for different relative payloads and steganalyzers.}
	\label{tab4}
	\begin{tabular}{ cl|cccc }
		\hline
		&&Chao&Li&VND/LSBR&HPQ-R\\
		\hline
		\multirow{4}{*}{LFS52} 
		& $\alpha=1.5$&$8.4556\times10^{-18}$&\textbf{1.000}&$8.4556\times10^{-18}$&$8.4556\times10^{-18}$\\
		&$\alpha=3$&$8.4556\times10^{-18}$&$0.0062$&$8.4556\times10^{-18}$&$8.4556\times10^{-18}$\\
		& $\alpha=4.5$ & $8.4556\times10^{-18}$&$7.7620\times10^{-10}$&$8.4556\times10^{-18}$&- \\
		& $\alpha=6$ & $8.4556\times10^{-18}$&\textbf{0.0892}&$8.4556\times10^{-18}$&- \\
		\hline
		\multirow{4}{*}{LFS64} 
		& $\alpha=1.5$&$8.4556\times10^{-18}$&$9.1907\times10^{-9}$&$8.4556\times10^{-18}$&$8.4556\times10^{-18}$\\
		&$\alpha=3$&$8.4556\times10^{-18}$&$4.2714\times10^{-7}$&$8.4556\times10^{-18}$&$8.4556\times10^{-18}$\\
		& $\alpha=4.5$ & $8.4556\times10^{-18}$&$3.5312\times10^{-9}$&$8.4556\times10^{-18}$&- \\
		& $\alpha=6$ & $8.4556\times10^{-18}$&$1.1071\times10^{-4}$&$8.4556\times10^{-18}$&- \\
		\hline
		\multirow{4}{*}{LFS76} 
		& $\alpha=1.5$&$8.4556\times10^{-18}$&$1.6945\times10^{-12}$&$8.4556\times10^{-18}$&$8.4556\times10^{-18}$\\
		&$\alpha=3$&$8.4556\times10^{-18}$&$1.6383\times10^{-8}$&$8.4556\times10^{-18}$&$8.4556\times10^{-18}$\\
		& $\alpha=4.5$ & $8.4556\times10^{-18}$&$2.5367\times10^{-17}$&$8.4556\times10^{-18}$&- \\
		& $\alpha=6$ & $8.4556\times10^{-18}$&$2.8411\times10^{-15}$&$8.4556\times10^{-18}$&- \\
		\hline
		\multirow{4}{*}{ELFS124} 
		& $\alpha=1.5$&$8.4556\times10^{-18}$&$3.2652\times10^{-5}$&$8.4556\times10^{-18}$&$8.4556\times10^{-18}$\\
		&$\alpha=3$&$8.4556\times10^{-18}$&$6.5049\times10^{-9}$&$8.4556\times10^{-18}$&$8.4556\times10^{-18}$\\
		& $\alpha=4.5$ & $8.4556\times10^{-18}$&$8.4556\times10^{-18}$&$8.4556\times10^{-18}$&- \\
		& $\alpha=6$ & $8.4556\times10^{-18}$&$1.6820\times10^{-10}$&$8.4556\times10^{-18}$&- \\
		\hline
		\multirow{4}{*}{NVT+} 
		& $\alpha=1.5$&$8.4556\times10^{-18}$&$6.9277\times10^{-14}$&$8.4556\times10^{-18}$&$8.4556\times10^{-18}$\\
		&$\alpha=3$&$8.4556\times10^{-18}$&$8.4556\times10^{-18}$&$8.4556\times10^{-18}$&$8.4556\times10^{-18}$\\
		& $\alpha=4.5$ & $8.4556\times10^{-18}$&$8.4556\times10^{-18}$&$8.4556\times10^{-18}$&- \\
		& $\alpha=6$ & $8.4556\times10^{-18}$&$8.4556\times10^{-18}$&$8.4556\times10^{-18}$&- \\
		\hline
       		\multirow{4}{*}{WFS228} 
		& $\alpha=1.5$&$8.4556\times10^{-18}$&$5.7154\times10^{-12}$&$8.4556\times10^{-18}$&$8.4556\times10^{-18}$\\
		&$\alpha=3$&$8.4556\times10^{-18}$&$8.4556\times10^{-18}$&$8.4556\times10^{-18}$&$8.4556\times10^{-18}$\\
		& $\alpha=4.5$ & $8.4556\times10^{-18}$&$8.4556\times10^{-18}$&$8.4556\times10^{-18}$&- \\
		& $\alpha=6$ & $8.4556\times10^{-18}$&$8.4556\times10^{-18}$&$8.4556\times10^{-18}$&- \\
		\hline
	\end{tabular}
\end{table*}
\bibliographystyle{IEEEtran}
\bibliography{mypaper}
\end{document}